\pdfoutput=1
\documentclass[usenatbib]{mnras}
\usepackage{hyperref}
\usepackage{amsfonts,amsmath,amssymb}
\usepackage{txfonts}
\usepackage{graphicx}
\usepackage{bm}
\usepackage{ctable}
\usepackage{url}
\usepackage{breakurl}
\usepackage{fixltx2e} 

\newcommand{\driftvel}{{\bf w}_{s}}
\newcommand{\driftvelmag}{w_{s}}
\newcommand{\driftveleq}{{\bf w}_{s,{\rm eq}}} 
\newcommand{\dustvel}{{\bf v}_{d}}
\newcommand{\gasvel}{{\bf u}_{g}}
\newcommand{\gasden}{\rho_{g}}
\newcommand{\dustden}{\rho_{d}}

\newcommand{\ts}{t_{s}}

\newcommand{\dug}{\delta u_{\text{gas}}}
\newcommand{\dud}{\delta v_{\text{dust}}}
\newcommand{\Dt}[1]{\frac{d #1}{dt}}
\newcommand{\grainsize}{\epsilon_{d}}
\newcommand{\grainsizedl}{\bar{\epsilon}_{d}}
\newcommand{\acceldl}{\bar{\rm a}}

\newcommand{\cs}{c_{s}}
\newcommand{\wavenumber}{k\,\langle \cs\rangle \langle\ts\rangle}
\newcommand{\kct}{k_{0}\,\langle c_{s}\rangle \langle\ts\rangle}
\newcommand{\magicparameter}{\mu/(k_{0}\langle\cs \rangle \langle\ts\rangle)}

\newcommand{\teddy}{t_{\text{eddy}}}
\newcommand{\internaldensity}{\rho_{d}^{i}}

\newcommand{\lowk}{\text{\small $\mu$0.01-$\acceldl$1e4-$\grainsizedl$0.001}}
\newcommand{\midk}{\text{\small $\mu$0.01-$\acceldl$100-$\grainsizedl$0.1}}
\newcommand{\highk}{\text{\small $\mu$0.1-$\acceldl$10-$\grainsizedl$1}}
\newcommand{\GG}[1]{} 
\newcommand{\addedtextkey}{}

\newcommand\altaffilmark[1]{$^{#1}$}
\newcommand\altaffiltext[1]{$^{#1}$}
\title[Nonlinear RDI]{Nonlinear Evolution of Instabilities Between Dust and Sound Waves\vspace{-0.5cm}}

\author[Moseley et al.]{
\parbox[t]{\textwidth}{ 
	Eric R.~Moseley\altaffilmark{1,2}\thanks{E-mail: moseley@princeton.edu}, Jonathan Squire\altaffilmark{1,3,4} \&\ Philip F.~Hopkins\altaffilmark{1,3}
} 
\vspace*{6pt} \\
\altaffiltext{1}{TAPIR, Mailcode 350-17, California Institute of Technology, Pasadena, CA 91125, USA} \\
\altaffiltext{2}{Department of Astrophysical Sciences, Princeton University, Princeton, NJ 08540, USA}\\
\altaffiltext{3}{Walter Burke Institute for Theoretical Physics, Pasadena, CA 91125, USA}\\
\altaffiltext{4}{Department of Physics, University of Otago, P.O.~Box 56, Dunedin 9054, New Zealand\vspace{-0.3cm}}
}

\date{Submitted to MNRAS, September, 2018\vspace{-0.6cm}}
\begin{document}
\maketitle

\begin{abstract}  
We study the non-linear evolution of the acoustic ``Resonant Drag Instability'' (RDI) using numerical simulations. The acoustic RDI is 
excited in a dust-gas mixture when dust grains stream through gas, interacting
with sound waves to cause a linear instability. We study this process in a periodic box by accelerating neutral dust with an external driving force. The instability grows as predicted by linear theory, eventually breaking into turbulence and saturating. As in linear theory, the non-linear behavior is characterized by three regimes -- high, intermediate, and low wavenumbers -- the boundary between which is determined by the dust-gas coupling strength and the dust-to-gas mass ratio. The high and intermediate wavenumber regimes behave similarly to one another, with large dust-to-gas ratio fluctuations while the gas remains largely incompressible. The saturated state is highly anisotropic: dust is concentrated in filaments, jets, or plumes along the direction of acceleration, with turbulent vortex-like structures rapidly forming and dissipating in the perpendicular directions. The low-wavenumber regime exhibits large fluctuations in gas and dust density, but the dust and gas remain more strongly coupled in coherent ``fronts'' perpendicular to the acceleration. These behaviors are qualitatively different from those of dust ``passively'' driven by external hydrodynamic turbulence, with no back-reaction force from dust onto gas. The virulent nature of these instabilities has interesting implications for dust-driven winds in a variety of astrophysical systems, including around cool-stars, in dusty torii around active-galactic-nuclei, and in and around giant molecular clouds. 
\end{abstract}

\begin{keywords}
instabilities --- turbulence --- ISM: kinematics and dynamics --- star formation: general --- galaxies: formation ---  planets and satellites: formation\vspace{-0.5cm} 
\end{keywords}

\vspace{-1.1cm}

\section{Introduction}
Since its discovery by \citet{trumpler1930spectrophotometric}, astrophysical dust has been recognized as important in nearly all areas of astronomy. In addition to its extinction and scattering effects, dust dynamics are important for star and planet formation  \citep{1996rdfs.conf.dust.star.formation,chiang:2010.planetesimal.formation.review}, the evolution of cool stars \citep{2012Natur.484..220N}, stellar and AGN ``feedback processes'' (e.g.\ radiation pressure-driven winds; \citealp{thompson:rad.pressure}), cooling in the ISM and protostellar disks, and chemical evolution \citep{draine:2003.dust.review}. Because dust is a collection of (often charged) aerodynamic particles that are imperfectly coupled to the gas, its dynamics  cannot be trivially related to the better understood gas dynamics.

In, for example, planetesimal formation -- perhaps the best-studied astrophysical application  where non-trivial dust dynamics play a crucial role -- the imperfect dust-gas coupling  produces
phenomena such as dust ``traps'' (in, e.g.,\ vortices and pressure bumps; \citealp{johansen2014multifaceted}), turbulent concentration of grains, and the ``streaming instability'' \citep{youdin.goodman:2005.streaming.instability.derivation}. It is increasingly believed that these dust-clustering mechanisms, especially the streaming instability, may resolve the decades-old problem of how to aggregate or grow grains from millimeter sizes through to planetesimals (see, e.g., \citealp{johansen2009particle, bai2010dynamics, yang2017concentrating}).

A recent series of papers, \citet{squire2018resonant,hopkins2018resonant,squire2018protoplanetary,hopkins2018ubiquitous},
have  demonstrated the existence of a generic super-class of instabilities that appear whenever dust moves through gas.
 These instabilities, termed ``Resonant Drag Instabilities'' (RDIs), generalize the streaming instability to a wide variety of astrophysical scenarios and systems, suggesting that dust-gas mixtures are usually unstable. 
 RDIs typically have growth rates which are maximized at ``resonant'' angles and wavenumbers, where the phase velocity of some wave in the underlying gas medium (e.g.,\ acoustic, magnetosonic, or Alfv\'en waves) matches that of the dust drift. In fact, a unique set of RDI sub-families appears for every possible ``resonant pair'' of dust and gas modes. For example, the ``streaming instability'' of \citet{youdin.goodman:2005.streaming.instability.derivation} can be understood to arise due to the resonance of dust drift and gas epicyclic modes.

The purpose of this paper is to move beyond the linear analyses of \citet{squire2018resonant,hopkins2018resonant, squire2018protoplanetary,hopkins2018ubiquitous}, and study the 
nonlinear regime of the RDI using numerical simulations. We consider one of the simplest  setups  possible: neutral dust, drifting under the influence of a constant driving force through a neutral (hydrodynamic), homogeneous gas medium. Physically, this situation can arise when the dust, but not the gas, is subject to an external force, such as radiation pressure.
The gas supports undamped sound waves, which can resonate with the drifting 
dust, destabilizing the ``acoustic RDI.'' 
 The linear regime of this instability was studied in detail in \citet{hopkins2018resonant}, but simulations are required to study the non-linear regime, including RDI-generated turbulence and orbit crossing in the dust. Although the setup is simple, it is important to understand the non-linear saturation of these instabilities because (1) they can provide fundamental insights into the behaviors of other, more complicated RDIs, and (2) the acoustic RDI may represent the fastest-growing RDI in many physical regimes and objects, including dusty cool-star (AGB or red giant) winds, dust in dense molecular clouds and cores, and the obscuring ``torus'' around AGN (see \citealt{hopkins2018ubiquitous} 
 for extensive discussion).

We emphasize that the setup and simulations in this work are fundamentally distinct from previous theoretical works that consider dust as a ``passive'' entity in an externally-turbulent medium \citep[e.g.][]{hogan1999scaling,cuzzi:2010.planetesimal.masses.from.turbulent.concentration.model,
hopkins2016fundamentally,lee:dynamics.charged.dust.gmcs}. In such studies, the gas is unaffected by the dust, \emph{viz.,} the momentum back-reaction of the dust  on the gas is ignored, an assumption that is truly valid only in the zero dust-to-gas mass ratio limit. 
In contrast, this dust back-reaction is a crucial ingredient in causing RDIs, and without it, 
 the ensuing behavior is qualitatively different.

%
\begin{figure*}
\begin{center}
\includegraphics[width=1\textwidth]{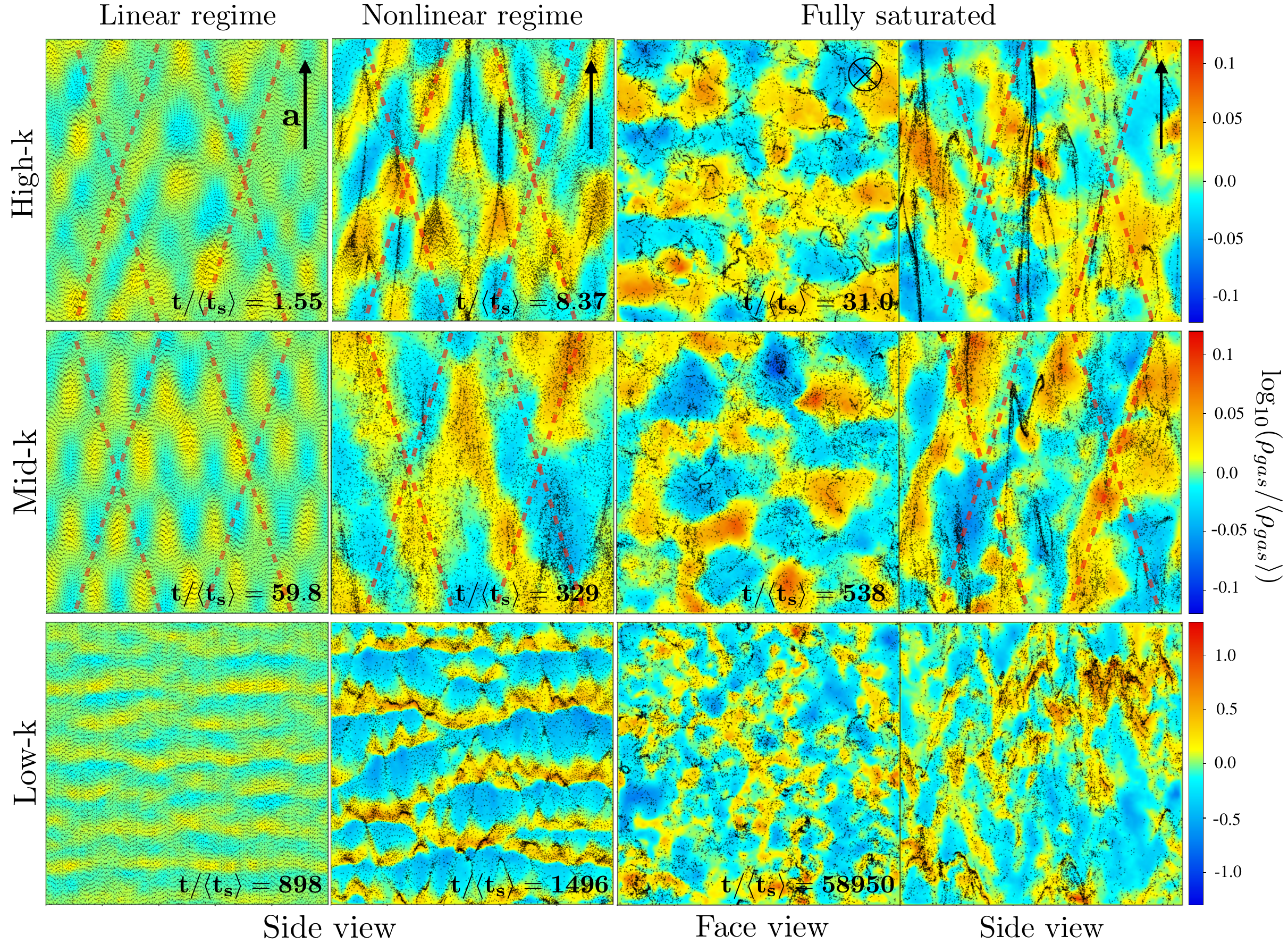}\vspace{-0.5cm}
\caption{Evolution of the acoustic RDI in three representative case studies (Tab.~\ref{tab:sims}) at high-$k$ ({\em top}), mid-$k$ ({\em middle}), and low-$k$ ({\em bottom}). Colors show gas density in a 2D slice through the 3D simulation box (color scale shown), with dust in slice shown as dots. Columns (1,\,2,\,4) show ``side-on'' slices (along direction of acceleration ${\bf a}$), while column (3) shows a ``face-on'' slice (perpendicular to ${\bf a}$); for clarity the direction of ${\bf a}$ is labeled in the upper right corner of each column. From left-to-right, panels show evolution into turbulence (times in units of equilibrium-state dust stopping time $\langle t_{s} \rangle$), illustrating (1) linear (or near-linear) initial evolution (modest dust \&\ gas fluctuations); (2) the early nonlinear regime (fluctuations are regular and resemble extreme linear instabilities); (3)-(4) fully non-linear regimes (with saturated box-scale turbulence). Dust and gas density fluctuations are less (more) strongly coupled on small/high-$k$ (large/low-$k$) scales. Dotted lines show the predicted wavefront orientation for the fastest-growing mode angle in linear theory. The modes visible in these figures are somewhat long wavelengths. This is because of the time we have chosen to show. It is sufficiently late that the smaller wavelength modes are no longer visible, as longer wavelengths have appreciably grown over them. Had we chosen an earlier time, the perturbations would not be visible with our chosen color scheme.}
\label{fig:regimes}
\end{center}
\end{figure*}
%
%

In \S~\ref{sec:methods}, we introduce the equations and numerical methods; \S~\ref{sec:theory} outlines some theoretical expectations. \S~\ref{sec:results} describes the results of our simulations, providing brief comparison to the theory. We conclude in \S~\ref{sec:conclusions}.

\section{Methods}\label{sec:methods}

As described above, our simulations are designed to study the basic physics of the acoustic RDI in the simplest 
setting possible. We thus consider a homogeneous mixture of dust and gas in a periodic box, 
with an external force acting on the dust only. Because of the drag between the gas and dust, this force accelerates  the gas as well as the dust, driving 
a mean velocity offset between the two phases. The box is thus simulating a small ``patch'' of a dust-gas mixture 
being accelerated by some external force (such as radiation pressure)  that acts differently upon the gas and dust.

\subsection{Equations Solved}

We directly integrate the equation of motion for a population of aerodynamic dust grains, each of which individually obeys
\begin{align}\label{eq:eom}
\Dt{\dustvel} &= -\frac{\driftvel}{t_{s}} + {\bf a},
\end{align}
where $\Dt{}$ is the Lagrangian (co-moving) derivative, ${\bf a}$ is the  external force/acceleration, $\ts$ is the drag coefficient or \textit{stopping time}, and $\driftvel \equiv \dustvel - \gasvel$ is the \textit{drift velocity}, defined as the difference between grain velocity ($\dustvel$) and gas velocity ($\gasvel$). We use angle brackets $\langle \cdot \rangle$ to denote a volume average. 

We will assume Epstein drag, with $t_{s}$ given by (see \citealt{draine.salpeter:ism.dust.dynamics})
\begin{align} \label{eq:tsdrift}
\ts(\gasden, \driftvel) &\equiv \sqrt[]{\frac{\pi \gamma}{8}}\frac{\internaldensity \,\grainsize}{\gasden\,c_{s}}\, \bigg( 1+\frac{9\pi\gamma}{128} \frac{\driftvel^{2}}{c_{s}^{2}} \bigg)^{-1/2},
\end{align}
where $\gasden$ is the gas density evaluated at the position  of the grain, $\gamma$ is the gas adiabatic index $\partial \log P/\partial\log\gasden$ (with $P$  the gas pressure), $\dustden^{i}$ is the \textit{internal} density of the grain, $\grainsize$ is the grain radius, and $\cs$ is the sound speed given by $\sqrt[]{\partial P/\partial \gasden}$. 
We denote the ratio of 
dust to gas mass density as $\mu \equiv \langle \dustden \rangle / \langle \gasden \rangle$. Technically, this expression is an approximation to the more general Epstein drag law, but the difference using a numerically-exact expression is completely negligible here \citep{draine.salpeter:ism.dust.dynamics}.

The gas obeys the usual Euler equations, but momentum conservation requires that we add the ``back-reaction'' term (drag force of grains on gas) to the momentum equation:
\begin{align}
\gasden\,\left(\frac{\partial}{\partial t} + \gasvel\cdot\nabla\right) \gasvel = -{\nabla P} + \int d^{3}\dustvel\,f_{d}(\dustvel)\,\frac{\driftvel}{t_{s}}.\label{eq:euler}
\end{align}
The latter term is simply the opposite of the force imparted by gas on grains, integrated over all dust grains at a given position. Here $f_{d}({\bf x},\,\dustvel)$ is the phase-space density distribution of dust --- i.e.\ differential mass of grains per element $d^{3}{\bf x}\,d^{3}\dustvel$ --- so the volumetric mass density of dust grains at a given position ${\bf x}$ is $\dustden ({\bf x}) \equiv \int d^{3}\dustvel f_{d}(\dustvel)$.

For this system, the grain properties are entirely specified by $\internaldensity\,\grainsize$ (the grain surface density). Because our primary goal is understanding the non-linear behavior of the instabilities, we will assume all grains in a given simulation have the same size, i.e.\ a single value of $\rho_{d}^{i}\,\grainsize$. In future work (in preparation) we will generalize to the more physical, but less easy-to-interpret, case of a general spectrum of grain sizes.

Critically, note that we {do not} make any fluid or ``local terminal velocity'' approximation for the dust, but integrate the trajectories of a population of grains directly. This is necessary to capture non-linear phenomena such as orbit crossings in the dust (otherwise, the non-linear outcomes would be unphysical).

\subsection{Numerical Methods}
\label{sec:numerical}

We solve equations \eqref{eq:eom}--\eqref{eq:euler} using the multi-method code {\small GIZMO} \citep{hopkins2014gizmo},\footnote{A public version of the code, including all methods used in this paper, is available at \href{http://www.tapir.caltech.edu/~phopkins/Site/GIZMO.html}{\url{http://www.tapir.caltech.edu/~phopkins/Site/GIZMO.html}}} using the second-order Lagrangian finite-volume ``meshless finite volume'' (MFV) method for the hydrodynamics, which has been well-tested on problems involving multi-fluid instabilities and shock-capturing. We model dust using the usual ``super-particle'' method \citep[e.g.][]{carballido2008kinematics,2007ApJ...662..613Y,bai:2010.grain.streaming.sims.test, pan2011turbulent}, whereby the motion of each ``dust particle'' in the simulation follows Eq.~\ref{eq:eom}, but each represents an ensemble of dust grains of size $\grainsize$ (in other words, we sample some finite, computationally feasible number of grains). The numerical methods for this integration are described and tested in \citet{hopkins2016fundamentally}. {The numerical implementaion of the back-reaction and other aspects of our methods are described in detail in Appendix ~\ref{numerics}.}


The gas and dust integration capabilities of {\small GIZMO} have been extensively  validated in \citet{hopkins2014gizmo} and \citet{hopkins2016fundamentally}. For this work, we have also run various resolution tests (see Fig.~\ref{fig:variance}) and compared to analytic solutions 
in the linear regime of the RDI.
In addition, we have run several simulations which are identical to others in our suite, except for numerical methods. Such tests include (1) a different hydrodynamic solver (the meshless-finite-mass or ``MFM'' method), (2) a different time integration and reconstruction scheme for the grains (a less accurate cell-centered scheme), (3) different initial conditions (using a glass-like, instead of regular, initial cell configuration), (4) different box geometries (boxes more extended in the drift direction by a factor $\sim 2-8$). As discussed in more detail in Appendix \ref{numerics}, these changes do not significantly affects our conclusions. We have also run a large suite of 2D simulations (reaching $\sim 2\times 2048^{2}$ resolution), which we do not present here, but produce qualitatively largely similar behavior in the dust, despite the (expected) different non-linear behavior of the induced gas turbulent cascade. 

Our default simulations adopt a resolution $N_{\rm gas} = 128^{3}$ gas elements, with an equal number of dust elements $N_{\rm dust}$. We have confirmed  our major results with higher resolution simulations at $N=256^{3}$, and in fact also find similar behavior in lower-resolution ($N=64^{3}$) simulations run for testing purposes. Additional numerical details of our methods, and code validation/convergence tests, are given in Appendix~\ref{numerics}.

\subsection{Equilibrium Solution and Initial Conditions}

In our simulations, only the dust is driven by the (constant) external radiation force (${\bf a}$), while the 
dust back-reaction on the gas acts to accelerate the gas. 
As discussed in detail \citet{hopkins2018resonant}, this leads to a quasi-equilibrium solution where the entire dust-gas system 
accelerates. Simultaneously, there is also a mean velocity offset between the dust and gas that balances the drag force against the radiation force and the system's own inertia. {Because the acceleration of the system as a whole is constant, the physics of the RDI in this accelerating reference frame is unchanged as compared to an intertial frame. That is to say, any fictitious force that emerges in the equations of motion appears in the equations for both dust \textit{and} gas, and thus does not affect the interplay between the two. To ensure numerical accuracy, we have checked this explicitly in our simulations by adding a constant external force to both the gas and dust to keep the mean velocity of the system as zero. For analysis purposes, we transform back into the reference frame where the system is stationary.}


In detail,
the quasi-equilibrium steady-state solution, from which our simulations are initialized, is homogeneous, with uniform mean dust-to-gas ratio $\mu \equiv \langle \dustden \rangle / \langle \gasden \rangle$, equilibrium gas velocity $\langle \gasvel \rangle$, and drift velocity $\langle \driftvel \rangle\equiv \langle \dustvel \rangle-\langle \gasvel \rangle$ { given by,}
\begin{align}
    \langle \gasvel \rangle &= \langle {\bf u} \rangle(t=0) + {\bf a}\,t\,\mu/(1+\mu)  \\
    \langle \driftvel \rangle &= {\bf a}\,\langle t_{s} \rangle / (1+\mu),
\end{align}
where $\langle t_{s} \rangle = t_{s}(\langle \gasden \rangle,\,\langle \driftvel \rangle)$ is the stopping time at the equilibrium drift velocity, and its  dependence of on $\driftvel$ implies that the equilibrium drift velocity depends on the detailed form of the Epstein drag law, Eq.~\eqref{eq:tsdrift}. From hereon, we shall denote this equilibrium value of $\driftvel$ as $\driftveleq$. This should be distinguished from the  value measured in the saturated state of a simulation $\langle \driftvel \rangle_{\rm sat}$, which can differ from $\driftveleq$ due to the RDI-generated turbulence. 

Our simulations begin from these equilibrium solutions at $t=0$: we initialize a 3D periodic (cubic) box of side-length $L_{0}$ with uniform dust and gas densities, $\langle {\bf u} \rangle(t=0) = 0$, and $ \dustvel = \driftvel  = \driftveleq = {\bf a}\,\langle t_{s} \rangle / (1+\mu)$ for all dust particles. 

\subsection{Units}

A convenient unit system for our simulations is formed using the equilibrium gas sound speed $\langle \cs \rangle$ and gas density $\langle \gasden \rangle$, and box size $L_{0}$. In these units, the initial conditions and outcome of the simulations is (at a given resolution) entirely determined by three dimensionless parameters: the mean dust-to-gas ratio $\mu$, the acceleration 
\begin{equation}
\acceldl\equiv{\bf a}\,L_{0}/\langle \cs\rangle^{2},\label{eq:dimless.a}
\end{equation}
and the grain size parameter\begin{equation}
\grainsizedl\equiv \internaldensity\,\grainsize / \langle \gasden \rangle \,L_{0}.\label{eq:dimless.grainsize}
\end{equation}
 From the equilibrium solutions above, we  see that these are equivalent to specifying the dimensionless equilibrium drift Mach number $\langle \driftvel \rangle / \langle \cs \rangle $, the stopping time $\langle t_{s} \rangle\,\langle \cs \rangle/L_{0}$, and $\mu$. For convenience, we define the smallest wavenumber that can fit in the box, $k_{0} \equiv 2\pi/L_{0}$. 

\subsection{Simulations}

Table~\ref{tab:sims} lists the ``production'' simulations that we have run as part of this study. In most  simulations, the equilibrium dust drift is supersonic ($\driftveleq/\langle \cs \rangle \gtrsim 1$). Although various different RDIs may be important even when the drift is highly subsonic (for example, the ``streaming instability'' and ``settling instability'' with gas epicyclic waves; \citealp{squire2018protoplanetary}), the acoustic RDI studied here grows fastest when the drift is supersonic, and we only expect it to dominate over other RDIs in this regime. 
Consequently, in our default simulations, we adopt an isothermal equation-of-state (EOS) for the gas ($\gamma=1$), which is appropriate for GMCs and dense, neutral gas where cooling times are short and  motions are often highly supersonic (and where dust charge, which couples grains  to magnetic fields, can often be neglected).\footnote{Because our default equation-of-state is isothermal, we note that $\cs =\langle \cs \rangle$ is constant, and we can use $\cs$ and $\langle \cs \rangle$ interchangeably.} 

{Because of its greater astrophysical interest, most simulations study the gas-dominated regime with $\mu<1$.
For completeness, we also run a number of dust-dominated ($\mu\geq 1$) simulations, but provide only 
a cursory analysis of these because they do not exhibit any obviously interesting differences in behavior compared to the $\mu<1$ cases.}

\subsection{Analysis}
\label{sec:statistics}

Unless otherwise noted, all statistics computed here are volume weighted. Gas statistics can be computed directly from {\small GIZMO}  output, using the fact that a gas resolution element $a$ has  volume $m_{\rm gas}^{a}/\rho_{\rm gas}^{a}$.
Dust statistics are computed in the same way, using  the the local dust density in the vicinity of each dust super-particle,  $\rho_{\rm dust}^{a}$, and the mass of the super-particle $m_{p}^{a} = M_{\rm dust}/N_{\rm dust}$ (where $M_{\rm dust} = \mu\,M_{\rm gas}$ is the total dust mass in the box). Because $\rho_{\rm dust}^{a}$ is not used directly in the {\small GIZMO} calculation, it is computed in post-processing using an SPH-like local kernel density estimator: $\rho_{\rm dust}^{a} = \sum_{b} m_{p}^{b}\,W({\bf x}_{a}-{\bf x}_{b},\,H_{a})$, where $W({\bf x}_{a}-{\bf x}_{b},\,H_{a})$ is the usual cubic spline kernel with radius of compact support $2\,H_{a}$ and $H_{a} = (m_{p}^{a}/\rho_{\rm dust}^{a})^{1/3}$ is the local mean inter-particle spacing.

All simulations are run well past saturation to confirm they have reached steady-state, and all ``saturated'' quantities plotted are time-averaged across all snapshots after saturation (the exact cut used makes little difference to the values).

%
\begin{table*}
\begin{center}
 \begin{tabular}{||c c c c c c c c c c||} 
 \hline
1 & 2 & 3 & 4 & 5 & 6 & 7 & 8 & 9 & 10 \\
 \hline
  Name  & $\mu\equiv \frac{\langle \dustden\rangle}{\langle \gasden \rangle}$ & $\acceldl \equiv \frac{| {\bf a}|\,L_{0}}{\langle c_s \rangle ^{2}}$ & $\grainsizedl \equiv\frac{\internaldensity\grainsize}{\langle \gasden \rangle L_{0}}$ & Regime & $\frac{\langle c_s \rangle \langle t_s\rangle_{0}}{L_{0}}$ &  $\frac{\langle \driftvelmag \rangle_{\rm sat}}{\langle c_s \rangle}$ & $\frac{\langle \delta {u}_{\rm gas}^{2}\rangle^{1/2}_{\rm sat}}{\langle c_s \rangle}$ & 
 $\sigma^{\rm sat}_{\log{\rho_{\rm gas}}}$ &
 $\sigma^{\rm sat}_{\log{\rho_{\rm dust}}}$   \\ [0.5ex] 
 \hline\hline
  \text{$\mu$0.001-$\acceldl$100-$\grainsizedl$0.1} & 0.001 &100 & 0.1 &  mid-$k$  & 0.034 & 3.5 & 0.051 & 0.00052 & 0.14 \\
  \text{$\mu$0.001-$\acceldl$1e3-$\grainsizedl$0.001} & 0.001 &1000  & 0.001 & mid-$k$   & 0.00060 & 0.84 & 0.19 & 0.074 & 0.21\\
\text{$\mu$0.001-$\acceldl$1e3-$\grainsizedl$0.01} & 0.001 & 1000 & 0.01 & mid-$k$   & 0.0034 & 3.7 & 0.34 & 0.042 & 0.39 \\
  \text{$\mu$0.001-$\acceldl$1e3-$\grainsizedl$0.1} & 0.001 &1000 & 0.1 & mid-$k$    & 0.011 & 12 & 0.13 & 0.030 & 0.19 \\
 \hline
 \text{ $\mu$0.01-$\acceldl$1-$\grainsizedl$0.1} & 0.01 & 1 & 0.1  &  mid-$k$  & 0.063 & 0.063 & 0.013 & 0.00030 & 0.11 \\
 \textcolor{gray}{\text{$\mu$0.01-$\acceldl$1-$\grainsizedl$1}} & 0.01 & 1 & 1  & mid-$k$  &   0.60 & 0.60 & 0.012 & 0.00027 & 0.046 \\
  \textcolor{gray}{\text{$\mu$0.01-$\acceldl$1-$\grainsizedl$10}} & 0.01 & 1 & 10 &  mid--high  & 3.3 & 3.3 & 0.013 & 0.00027 & 0.045 \\
 \text{$\mu$0.01-$\acceldl$10-$\grainsizedl$0.1} & 0.01 & 10 & 0.1  & mid-$k$  & 0.060 & 0.60 & 0.013 & 0.00027 & 0.048 \\
 \text{$\mu$0.01-$\acceldl$10-$\grainsizedl$1} & 0.01 & 10 & 1 &  mid-$k$  & 0.33 & 3.3 & 0.097 & 0.0030 & 0.31 \\
 \text{$\mu$0.01-$\acceldl$100-$\grainsizedl$0.01} & 0.01 & 100 & 0.01  &  mid-$k$ & 0.0060 & 0.73 & 0.12 & 0.034 & 0.13 \\
 \textit{$\mu$0.01-$\acceldl$100-$\grainsizedl$0.1-LR} & 0.01 & 100 & 0.1 & mid-$k$ & 0.033 & 3.3 & 0.35 & 0.028 & 0.071 \\
 \textbf{$\mu$0.01-$\acceldl$100-$\grainsizedl$0.1} & \bf{0.01} & \bf{100} & \bf{0.1} &  \bf{mid-$\bm{k}$} & \bf{0.033} & \bf{3.3} & \bf{0.33} & \bf{0.025} & \bf{0.37} \\
 \textit{$\mu$0.01-$\acceldl$100-$\grainsizedl$0.1-HR} & 0.01 & 100 & 0.1 &  mid-$k$  & 0.033 & 3.3 & 0.32 & 0.033 & 0.43 \\
 \text{$\mu$0.01-$\acceldl$1e3-$\grainsizedl$0.001} & 0.01 & 1000 & 0.001  &  low--mid & 0.00060 & 0.85 & 0.53 & 0.14 & 0.23 \\
 \text{$\mu$0.01-$\acceldl$1e3-$\grainsizedl$0.1} & 0.01 & 1000 & 0.1 &  mid-$k$  & 0.011 & 11.8 & 0.88 & 0.063 & 0.37 \\
 \text{$\mu$0.01-$\acceldl$1e3-$\grainsizedl$0.1-$\gamma$5/3} & 0.01 & 1000 & 0.1 & mid-$k$   & 0.012 & 10 & 0.69 & 0.032 & 0.26 \\
 \text{$\mu$0.01-$\acceldl$1e3-$\grainsizedl$1} & 0.01 & 1000 & 1 &  mid-$k$  & 0.036 & 37 & 0.51 & 0.024 & 0.17 \\
 \textit{$\mu$0.01-$\acceldl$1e4-$\grainsizedl$0.001-LR} & 0.01 & 10000 & 0.001 & low-$k$ &  0.00033 & 4.6 & 3.5 & 0.20 & 0.33 \\
 \textbf{$\mu$0.01-$\acceldl$1e4-$\grainsizedl$0.001} & \bf{0.01} & \bf{10000} & \bf{0.001} & \bf{low-$\bm{k}$} &  \bf{0.00033} & \bf{4.6} & \bf{3.4} & \bf{0.22} & \bf{0.40}\\
 \textit{$\mu$0.01-$\acceldl$1e4-$\grainsizedl$0.001-HR} & 0.01 & 10000 & 0.001 &   low-$k$ & 0.00033 & 4.2 & 3.0 & 0.23 & 0.43 \\
 \text{$\mu$0.01-$\acceldl$1e4-$\grainsizedl$0.1} & 0.01 & 10000 & 0.1 &  mid-$k$ & 0.0036 & 37 & 20 & 0.19 & 0.50 \\
 \text{$\mu$0.01-$\acceldl$1e5-$\grainsizedl$0.1} & 0.01 & 100000 & 0.1 &  low--mid  & 0.0011 & 100 & 127 & 0.32 & 0.55 \\
 \text{$\mu$0.01-$\acceldl$1e6-$\grainsizedl$0.001} & 0.01 & 1000000 & 0.001 & low-$k$  & 0.000036 & 44 & 94 & 0.29 & 0.40 \\
  \hline
 \textcolor{gray}{\text{$\mu$0.1-$\acceldl$1-$\grainsizedl$10}} & 0.1 & 1 & 10  & high-$k$  & 3.2 & 3.0 & 0.013 & 0.00027 & 0.046 \\
 \text{$\mu$0.1-$\acceldl$10-$\grainsizedl$0.1} & 0.1 & 10 & 0.1  &  mid-$k$ & 0.060 & 0.55 & 0.013 & 0.00027 & 0.050 \\
 \textbf{$\mu$0.1-$\acceldl$10-$\grainsizedl$1} & \bf{0.1} & \bf{10} & \bf{1}  &  \bf{mid--high} & \bf{0.32} & \bf{3.2} & \bf{0.33} & \bf{0.021} & \bf{0.46} \\
 \text{$\mu$0.1-$\acceldl$10-$\grainsizedl$10} & 0.1 & 10 & 10 &  mid--high  & 1.1 & 2.2 & 0.76 & 0.047 & 0.32 \\
  \text{$\mu$0.1-$\acceldl$100-$\grainsizedl$0.1} & 0.1 & 100 & 0.1  & mid-$k$  & 0.032 & 3.2 & 1.6 & 0.11 & 0.45 \\
  \text{$\mu$0.1-$\acceldl$1e3-$\grainsizedl$0.001} & 0.1 & 1000 & 0.001  & low-$k$ & 0.00060 & 0.86 & 1.8 & 0.16 & 0.20 \\
  \text{$\mu$0.1-$\acceldl$1e3-$\grainsizedl$0.1} & 0.1 & 1000 & 0.1  &  low--mid & 0.011 & 2.4 & 32 & 0.17 & 0.51 \\
 \hline
  \text{$\mu$1-$\acceldl$10-$\grainsizedl$0.1} & 1 & 10 & 0.1  &    & 0.056 & 0.37 & 1.1 & 0.079 & 0.41 \\
  \text{$\mu$1-$\acceldl$100-$\grainsizedl$0.1} & 1 & 100 & 0.1  &   & 0.025 & 1.9 & 8.2 & 0.12 & 0.43 \\
  \text{$\mu$1-$\acceldl$1e3-$\grainsizedl$0.001} & 1 & 1000 & 0.001  &   & 0.00056 & 0.54 & 6.9 & 0.24 & 0.26 \\
  \text{$\mu$1-$\acceldl$1e3-$\grainsizedl$0.1} & 1 & 1000 & 0.1  &   & 0.0081 & 11 & 58 & 0.22 & 0.86 \\
 \hline
  \text{$\mu$10-$\acceldl$100-$\grainsizedl$0.1} & 10 & 100 & 0.1  &   & 0.011 & 1.1 & 6.1 & 0.1 & 0.33 \\
  \text{$\mu$10-$\acceldl$1e3-$\grainsizedl$0.001} & 10 & 1000 & 0.001  &   & 0.00032 & 0.14 & 1.4 & 0.16 & 0.15 \\
  \text{$\mu$10-$\acceldl$1e3-$\grainsizedl$0.1} & 10 & 1000 & 0.1  &   & 0.0034 & 3.6 & 15 & 0.18 & 0.38 \\
 \hline
  \text{$\mu$100-$\acceldl$10-$\grainsizedl$0.1} & 100 & 10 & 0.1  &   & 0.011 & 0.13 & 0.24 & 0.025 & 0.13 \\
  \text{$\mu$100-$\acceldl$100-$\grainsizedl$0.1} & 100 & 100 & 0.1  &   & 0.0036 & 0.19 & 1.1 & 0.15 & 0.27 \\
  \text{$\mu$100-$\acceldl$1e3-$\grainsizedl$0.1} & 100 & 1000 & 0.1  &   & 0.0011 & 0.86 & 1.6 & 0.20 & 0.20 \\
\hline
 \end{tabular}
\end{center}
\caption{Simulations studied here. We list: 
(1) Simulation name. 
(2) Total dust-to-gas mass ratio $\mu$. 
(3) Differential acceleration $\bf{a}$ imposed between dust and gas (in dimensionless units of box size $L_{0}$ and isothermal sound speed $\langle c_{s} \rangle$). 
(4) Dimensionless grain size/surface density parameter $\grainsizedl$, which determines the strength of the drag forces. 
(5) Regime of the simulation (low, mid, or high-$k$, or a mix), calculated from the 6 {longest wavelength}  modes in the box for simulations with $\mu<1$.
(6) Initial stopping time (drag coefficient) $t_{s}$ of grains in the homogeneous equilibrium setup: the ratio $c_{s} t_{s}/L_{0}$ defines the ``low''-$k$ vs.\ ``mid''/``high''-$k$ regimes.
 (7) Mean drift velocity of grains relative to gas $\langle w_{s} \rangle$, measured in the saturated turbulent state -- especially large values indicate systems where grains ``draft'' through narrow channels carved in the gas. 
 (8) Volume-weighted standard-deviation of gas velocity in the box, measured in the saturated state. 
 (9) Volume-weighted standard-deviation of logarithmic gas density ($\log_{10}(\gasden/\langle \gasden\rangle)$), again measured in the saturated state. 
 (10) Same for dust density. 
 Bold names denote the ``case studies'' used in Figs.~\ref{fig:regimes}-\ref{fig:du}. Gray names indicate that the simulation is likely affected by the numerical issues discussed in App.~\ref{sec:laminar}.
}
\label{tab:sims}
\end{table*}
%

%
%
\begin{figure}
\begin{center}
\includegraphics[width=1\columnwidth]{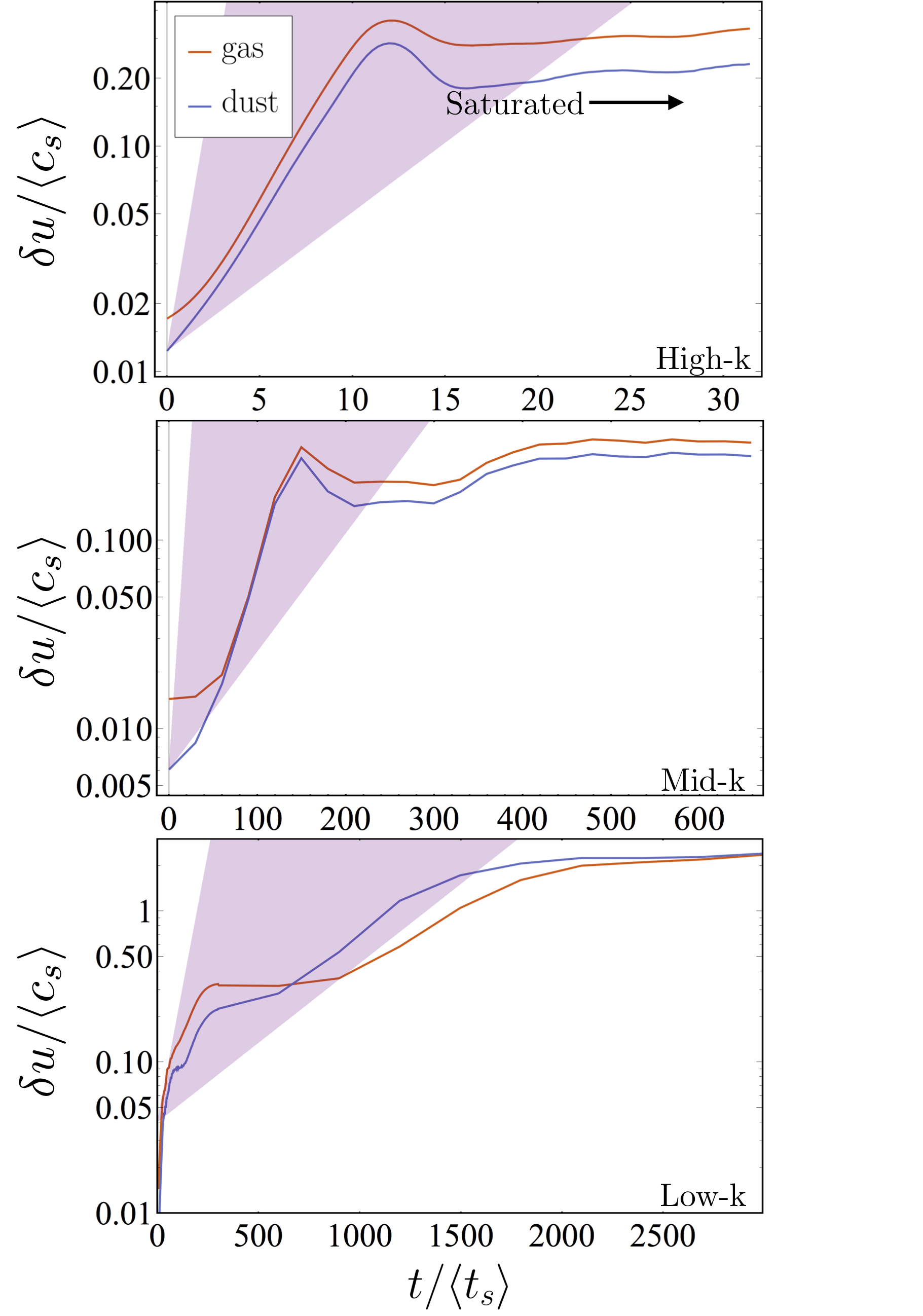}
\vspace{-0.5cm}
\caption{Time evolution of velocity dispersion in dust and gas versus time (in units of equilibrium stopping time $\langle \ts\rangle$). Panels show high/mid/low-$k$ representative ``case study'' simulations as Fig.~\ref{fig:regimes}. In each, the shaded fan shows the range of growth rates from linear theory (Eq.~\ref{eq:growthrates}), from wavenumbers $k=(1-128)\,2\pi/L_{0}$ (where $N=128$ is the linear resolution), which is consistent with the early growth rates. Statistics of the turbulence are measured in the saturated states as identified.}
\label{fig:evolution}
\end{center}
\end{figure}
%
%

%
\begin{figure*}
\begin{center}
\includegraphics[width=1.0\textwidth]{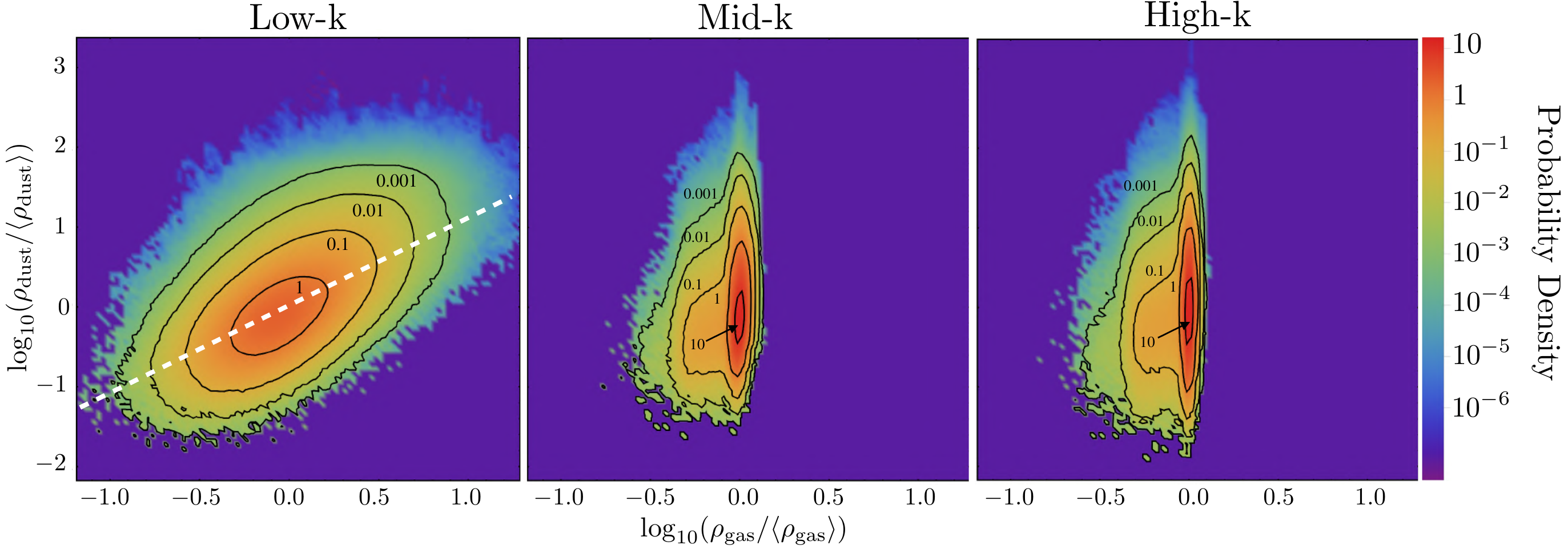}\vspace{-0.5cm}
\caption{Volume-weighted joint probability density functions (PDF) of gas and dust density for the simulations in Fig.~\ref{fig:regimes}. At low-$k$, dust and gas are relatively well-correlated. {The white dashed line in this panel is where dust and gas fluctuate exactly together, $\rho_d \propto \rho_g$.} At mid/high-$k$, an asymmetric PDF develops where gas develops a long tail at low densities (characteristic of subsonic turbulence), while dust remains relatively symmetric with a much larger density dispersion.
 }
\label{fig:jointhist}
\end{center}
\end{figure*}
%
%

\section{Theoretical Expectations}\label{sec:theory} 

\subsection{Linear Theory: The Three Regimes}\label{sec:linear.theory} 

\citet{hopkins2018resonant} considered a linear stability analysis of the equations solved here, adopting the usual Fourier decomposition of perturbations $\delta f \propto \exp{[i\,({\bf k}\cdot{\bf x} - \omega\,t)]}$ (for some field $f$, such as $\gasvel$ or $\gasden$). They showed that the behavior of the interesting unstable solutions depended critically on the dimensionless wavenumber parameter $\wavenumber \equiv |{\bf k}|\, c_{s} \langle t_{s} \rangle$ of the mode. Three regimes emerge:
\begin{equation}\label{eq:regimes}
\begin{cases}
\ \wavenumber \lesssim \mu \hspace{0.1cm}&\text{(Low-$k$)},\\
\ \mu \lesssim \wavenumber \lesssim \mu^{-1}\hspace{0.1cm}&\text{(Mid-$k$)},\\
\ \wavenumber \gtrsim \mu^{-1}\hspace{0.1cm}&\text{(High-$k$)}.
\end{cases}
\end{equation}
The linear behavior of the mid-$k$ and high-$k$ regimes is qualitatively similar, albeit with somewhat different scalings for the growth rates and mode structure. In both cases the fastest-growing modes are those that are ``resonant,'' with $\hat{\bf k} \cdot \langle \driftvel \rangle = \pm \cs$, i.e.\ with the drift velocity in the direction of the mode matching the sound speed. We can also see above that as $\mu \rightarrow 0$, the mid-$k$ regime extends to all $k$; it is therefore also the ``low-$\mu$'' regime. 

In the low-$k$ regime, in contrast, the fastest-growing mode is the so-called ``pressure-free'' mode. In this regime, the bulk force from dust on gas becomes larger than pressure forces, so the gas becomes highly compressible, and the fastest-growing modes are those with $\hat{\bf k}$ aligned with ${\bf a}$ (or $\langle \driftvel \rangle$). 

{If $\mu \gtrsim 1$, the definition of these regimes deteriorates, because the idea of the 
resonance between dust and sound waves  becomes ill defined since sound waves are strongly 
damped. However, the linear mode structure of the low-$k$ and high-$k$ regimes remains broadly 
unchanged \citep{hopkins2018resonant}, and our simulations at $\mu\geq1$ appear broadly similar to the $\mu<1$ simulations. Because we lack a theory for  the physics of the high-$\mu$ regime, and given its lesser physical interest, we  provide only cursory analysis of $\mu\geq 1$ simulations
 and do not consider it in the simple analytic scalings derived below.}

{These regimes also apply formally only to individual wavenumber parameters $k \langle c_s \rangle \langle t_s \rangle$. As a result, a simulation that spans a range of wavenumbers may technically fall into more than one regime. For ease of discussion, we have adopted the convention that the regime a simulation falls into (cf. Tab.~\ref{tab:sims}, col. 5) is that which the six longest wavelength modes $k \in [k_0,6 k_0]$ fall into (where $k_0 \equiv 2 \pi/L_0$, with $L_0$ being the size of the simulation box). Note that our choice to use the six longest wavelength modes is somewhat arbitrary. If some of these modes are in one regime while the rest are in another, they are classified as being in both regimes (e.g.  Tab.~\ref{tab:sims}, row 26).
}

{The definition of the regimes given in the expression \eqref{eq:regimes} suggests that the  parameter $\magicparameter$ may be useful for predicting and discussing the behavior of simulations.  That the behavior of a simulation would change with this parameter can be understood intuitively by noting that $\magicparameter$ is the ratio of the force that dust drifting at $\driftvelmag\sim c_s$ exerts on the gas to the pressure forces on the gas (for a mode of scale $k$). We will thus sometimes term $\magicparameter$ as the ``force parameter.'' As the mass of dust goes up or down, so does this force; similarly, at smaller scales, the pressure forces become more significant thus decreasing $\magicparameter$. The mid-$k$ and low-$k$ regimes are separated by $\magicparameter = 1$.}

The linear growth rates of the fastest-growing modes in each regime are approximately given by \citep{hopkins2018resonant} ,
\begin{equation}\label{eq:growthrates}
\Im(\omega)\, \langle \ts \rangle \sim \begin{cases}
\ \mu^{1/3}\,(\wavenumber)^{2/3}\,(\langle \driftvelmag \rangle /\cs)^{2/3} &\text{(Low-$k$)},\\
\ \mu^{1/2}\,(\wavenumber)^{1/2}\quad \text{at }\langle \driftvelmag \rangle>\cs &\text{(Mid-$k$)},\\
\ \mu^{1/3}\,(\wavenumber)^{1/3}\quad \text{at }\langle \driftvelmag \rangle>\cs &\text{(High-$k$)},
\end{cases}
\end{equation} 
where here $\langle \driftvelmag \rangle$ and other quantities refer, of course, to their values in the equilibrium, homogeneous solution. In Eqs.~\eqref{eq:regimes}-\eqref{eq:growthrates} we simplified by taking $\mu \ll 1$, which is chosen for most of the cases we consider here due to its greater astrophysical relevance. 
Growth rates for subsonic drift ($\langle\driftvelmag\rangle<\cs$) in the mid- and high-$k$ regimes are significantly lower and depend on  details of the equation of state and dust drag \citep{hopkins2018resonant}. From Eqs.~\eqref{eq:dimless.a}--\eqref{eq:dimless.grainsize} or Tab.~\ref{tab:sims} (see column 6, which lists $\langle c_s\rangle \langle \ts \rangle/L_0$), we 
see that simulations with smaller (larger) $\grainsizedl$ or larger (smaller) $\acceldl$ will have smaller (larger) dimensionless wavenumbers $\wavenumber$.

\subsection{Expected Turbulent Scalings}

%
%
\begin{figure}
\begin{center}
\includegraphics[width=0.45\textwidth]{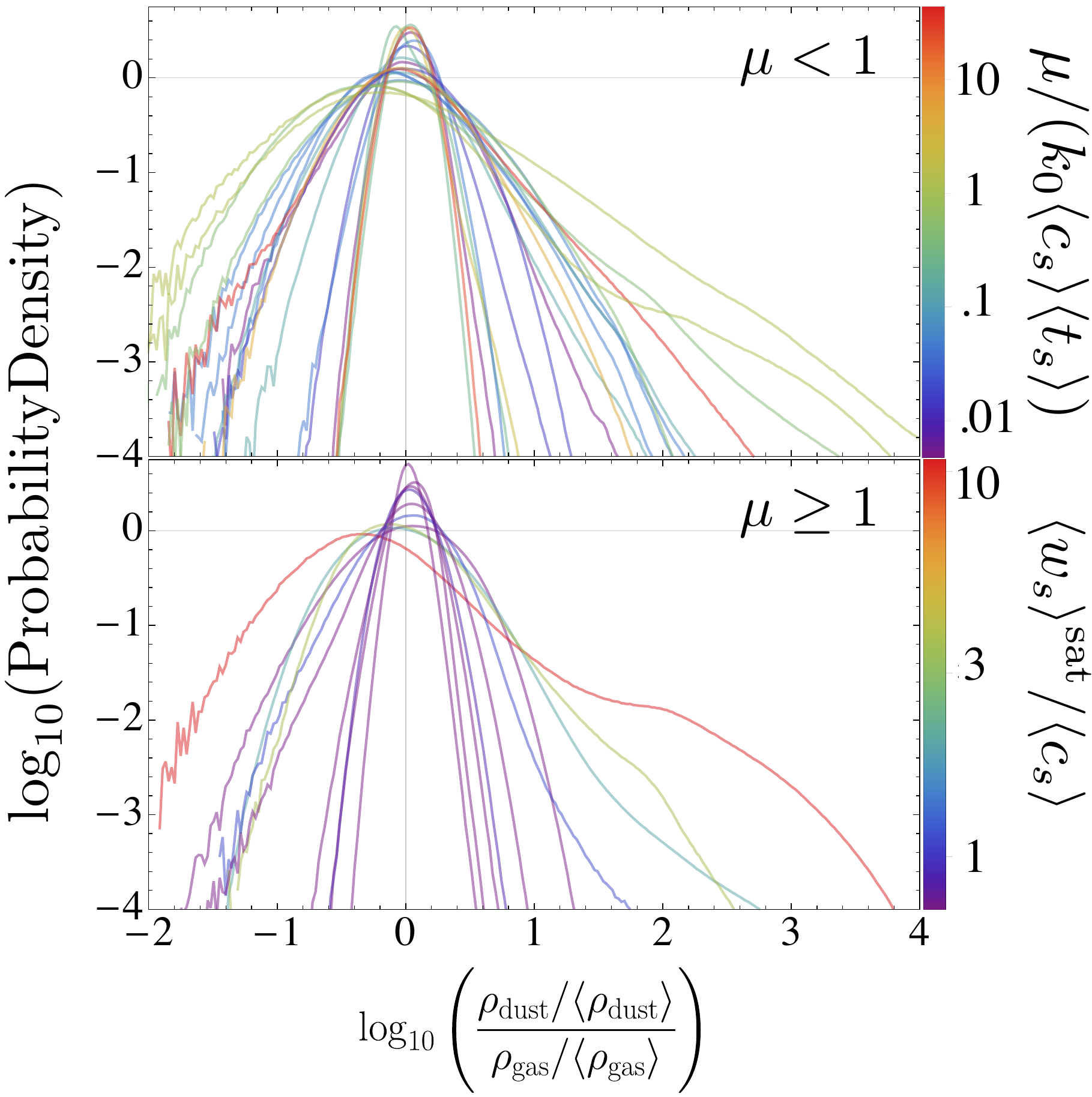}\vspace{-0.3cm}
\caption{{PDFs of the ratio of dust density to gas density (normalized) for all simulations, separated 
into simulations with $\mu<1$ (upper panel), and simulations with $\mu\geq 1$ (lower panel).
We color $\mu<1$ simulations according to the ``force parameter'' $\mu/(k_{0}\,\langle c_{s} \rangle \langle t_{s} \rangle)$, as discussed in \S\ref{sec:linear.theory}. $\mu \ge 1$ simulations are colored according to the dust drift velocity in the saturated state, $\langle w_s \rangle ^{\rm sat}/\langle \cs \rangle$, as the force parameter is not necessarily as physically meaningful for this regime.
There is some trend towards larger $\dustden/\gasden$ fluctuations at larger $\mu/(k_{0}\,\langle c_{s} \rangle \langle t_{s} \rangle)$  (low-$k$ regime), although there 
is significant variation around this. This should be expected because
similar $\dustden/\gasden$ fluctuations can arise 
either through large local dust fluctuations in a relatively quiescent gas (for 
simulations in the mid- and high-$k$ regimes), or through correlated 
gas and dust density fluctuations in more vigorous turbulence (in the low-$k$ regime). See also   Fig.~\ref{fig:jointhist} and the bottom-right panel of Fig.~\ref{fig:scaling}.
The PDFs mostly have log-normal-like shapes, although generally with  significant
high-density tails. Some distinctly non-Gaussian features (for instance, bumps) occur, particularly
in the high-$\mu$ regime, due to the dominance of individual large-scale structures in the saturated state.}}
\label{fig:pdf}
\end{center}
\end{figure}
%
%

Here we give simple, quasi-linear estimates for the amplitude of the turbulence driven by the acoustic RDI. We see below (\S~\ref{sec:results})  that these match the measured saturated states of the simulations relatively well.

The basic idea of the argument is to match the turnover time of the turbulence on the largest scale in the box to the growth rate of the RDI.
Given the predicted growth rates $\Im(\omega)$, 
and assuming that the saturated state is turbulent, then when the turnover time of the largest eddies is smaller than their growth time $1/\Im(\omega)$, all scales in the box are mixed before they can grow. Thus, a reasonable estimate for the saturation amplitude of  RDI generated turbulence is when $\teddy^{-1} \sim k_{0}\,\dug \sim k_{0}\,\dud \sim \Im(\omega)$, giving,
\begin{equation}\label{eq:duscaling}
\frac{\dug}{\cs} \sim \begin{cases}
	\ \mu^{1/3}\, (\langle \driftvelmag \rangle /\cs)^{2/3}\, (\kct)^{-1/3} & (\text{Low-$k$})\\
    \ \mu^{1/2}\, (\kct)^{-1/2} & (\text{Mid-$k$})\\
    \ \mu^{1/3}\, (\kct)^{-2/3} & (\text{High-$k$}).
\end{cases}
\end{equation}
Combining the regime definitions (Eq.~\ref{eq:regimes}) with these expressions, we see that the mid-$k$ and high-$k$ regimes can generate only subsonic gas turbulence when $\mu < 1$, with $\mu \lesssim \dug/\cs \lesssim 1$ in the  mid-$k$ regime, and $\dug/\cs \lesssim \mu$ in the high-$k$ regime. In contrast, turbulence generated by the  low-$k$ mode can be supersonic in the gas  for sufficiently low $k$ or high $\langle \driftvelmag \rangle/\cs$ (with $\dug/\cs \gtrsim (\langle \driftvelmag \rangle/\cs)^{2/3}$).

From this, we can speculate further about gas density statistics. In isothermal turbulence, the gas develops an approximately log-normal density distribution with variance,
\begin{equation}\label{eq:dens.var.mach.num}
    \sigma^{2}(\log_{10}\rho) \approx (\ln{10})^{-2}\,\ln{[1 + (b\,\dug/\cs)^{2}]},
\end{equation} where $b\sim 1/4-1$, depending on the forcing \citep[e.g.][]{federrath:2008.density.pdf.vs.forcingtype}. From the above (and assuming $b\sim 1/3$), we  obtain, {for each of the regimes,}
\begin{equation}
\sigma(\log_{10} \rho) \approx \begin{cases}
0.1(\delta u_{\rm gas}/\cs) & (\text{Mid-}/\text{High-$k$}) \\
0.43 \ln[1+\frac{1}{9}(\delta u_{\text{gas}}/c_s)^{2}]^{1/2} & (\text{Low-$k$}).
\end{cases}
\end{equation}
{For the mid-/high-$k$ regimes, $\sigma(\log_{10} \rho) \ll 1$ always, while in the low-$k$ regime, $\sigma(\log_{10} \rho)$ can be larger than 1, but because of the square-root/logarithmic suppression should lie in the range $\sim 0.1-0.6$ for all parameters simulated here.}

{In the saturated state for the low- and mid-$k$ regimes of the RDI (though not necessarily the high-$k$ regime), the dust stopping length is shorter than the size of the largest gas eddies. Given this,}
it is reasonable to assume that dust and gas are well mixed and efficiently share energy locally with respect to the local mean drift. This implies that that two should have similar velocity dispersions,  $\dud\sim\dug$. 
However, the corresponding dust density fluctuations are more difficult to predict, because the dust is pressure-free, meaning its density fluctuations need not be linked straightforwardly to its velocity dispersion. A detailed theory of dust fluctuations in saturated RDI turbulence will be explored in future work.

\section{Results}\label{sec:results}

\subsection{Representative Case Studies}\label{sub:representative.case.studies}

We find that the three regimes from linear theory  extend to describe many features of the nonlinear saturation as well, so we frame our discussion around these. Every simulation box contains a range of wavenumbers, of course, but given our finite resolution, it is typically the case that most of the resolved modes with $2\pi/L_{0} \lesssim k \lesssim 1/\Delta x$ (where $\Delta x \equiv L_{0}/N^{1/3}$ is the grid scale) lie in one particular regime for a given run. We therefore identify one simulation primarily within each regime (\lowk, \midk, and \highk) as representative of those in our parameter survey (although the longest wavelength modes of \highk\ lie in the mid-$k$ regime, due to numerical difficulties in capturing the high-$k$ instability; see App.~\ref{sec:laminar}).

%
\begin{figure}
    \centering
    \includegraphics[width=\columnwidth]{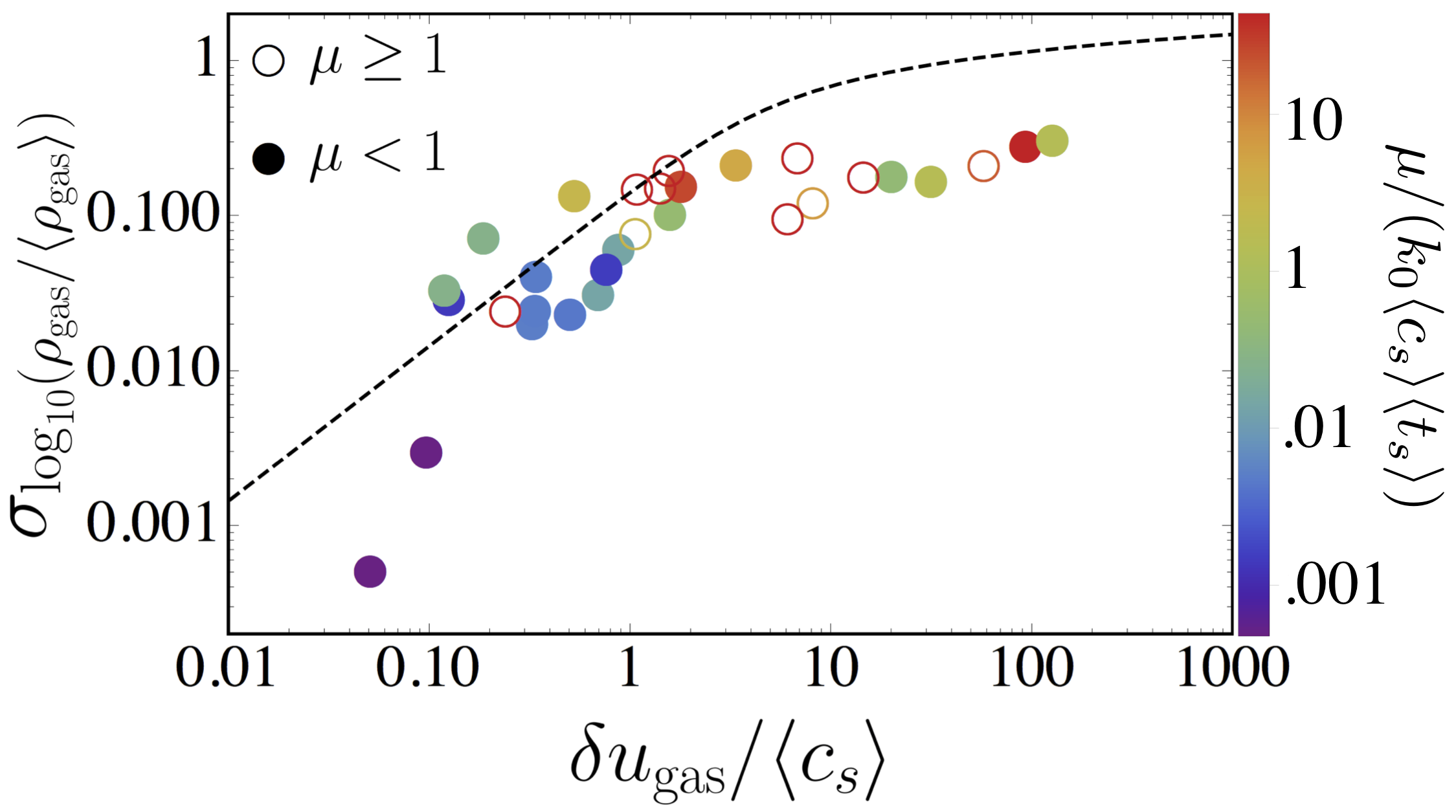}\vspace{-0.5cm}
    \caption{Standard deviation of logarithmic gas density, versus gas velocity dispersion. Plotted line is the standard scaling from \citet{federrath:2008.density.pdf.vs.forcingtype} with $b = 1/3$. This provides a reasonable fit for $\delta u_{\rm gas} \ll 3\,c_{s}$, but the simulations here produce weaker density fluctuations when $\delta u_{\rm gas} \gtrsim 3 \,c_{s}$, as these extreme cases tend to involve gas moving rapidly with dust along ``channels'' (rather than e.g.\ isotropic compressible turbulence in the gas).}
       \label{fig:density_scaling}
\end{figure}
%

%
%
\begin{figure}
\begin{center}
\includegraphics[width=0.45\textwidth]{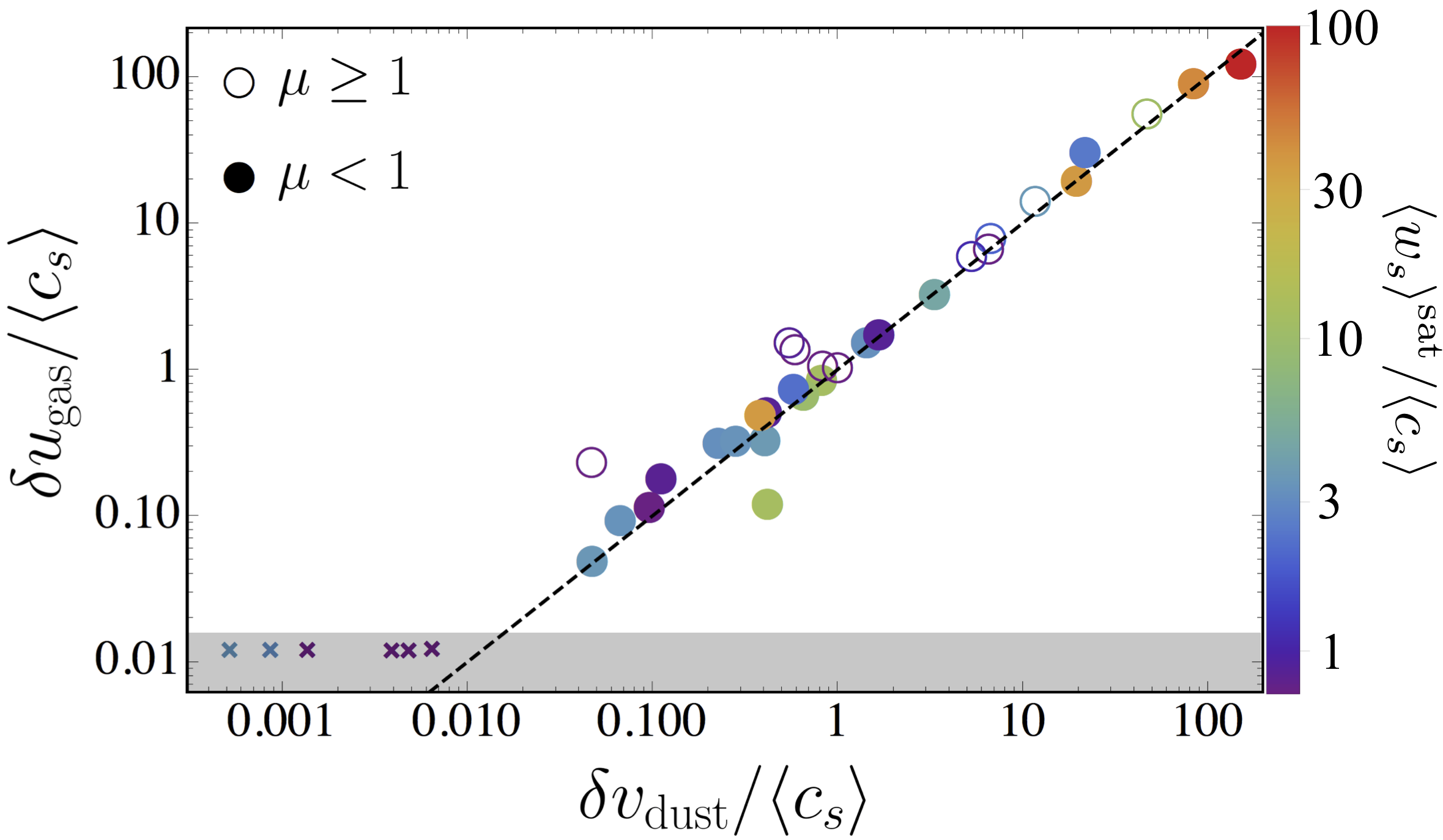}\vspace{-0.3cm}
\caption{Gas vs.\ dust velocity dispersions, with the dashed black line being equivalence. In saturation, the two are comparable with relatively small scatter. In most cases, the mean/bulk drift velocity $\langle \driftvelmag \rangle / \langle \cs \rangle$ of the saturated state (colors, as labeled) does not correlate strongly with the {\em fluctuations} $\delta u_{\rm gas}$, $\delta v_{\rm dust}$ in either dust or gas. The most extreme cases, however, do have large fluctuations {\em and} large drift velocities. The six simulations marked by ``x''s in the bottom left failed to go turbulent due to reasons discussed in \S~\ref{sec:laminar}. They reside within a grayed out, numerically limited region. They are also listed and grayed out in Tab. ~\ref{tab:sims}.}
\label{fig:du}
\end{center}
\end{figure}

\subsubsection{Linear growth of the instability}

At early times, the acoustic RDI behaves as predicted in \citet{hopkins2018resonant}. As seen in the first column of Figure~\ref{fig:regimes}, the instability begins as sinusoidal oscillations, with the dominant (fastest-growing) wavevector clearly aligned along a characteristic angle. For the mid-$k$ and high-$k$ modes this is the predicted ``resonant angle'' ($\hat{\bf k}\cdot \langle \driftvel \rangle = \pm \cs$), while for the low-$k$ this is   aligned with the drift $\hat{\bf k}\, \| \,\langle \driftvel \rangle$, as expected.

Further, as seen in Figure~\ref{fig:evolution}, the growth rate of the instabilities at early times agrees well with the linear theory predictions assuming a fastest-growing wavenumber $k_{\rm max}$ between $\sim 1/\Delta x$ and $\sim 3/\Delta x$. Note that because the predicted growth rates of this instability increase without limit with $k$, the fastest-growing mode will always be the {highest wavenumber} resolved. This is not exactly defined, but occurs at some multiple of the grid scale. The gas velocity and density fluctuations in the mid- and high-$k$ cases grow at approximately this rate until saturation, which is perhaps not surprising since they saturate in a quasi-linear regime. In the low-$k$ case the growth rates slow down, but do not vanish, as the perturbations become more strongly non-linear ($\dug/\cs \rightarrow 1$). 

In all cases the dust density perturbations, which do not incur any restoring pressure forces, become strongly non-linear well before the gas. This can be clearly seen in the second column of Figure~\ref{fig:regimes}. 

Note that in certain simulations at high $k$ and low $\mu$ -- e.g. $\mu$0.01-$\acceldl$1-$\grainsizedl$10 -- the instability does not grow as expected from linear theory, which we attribute to numerical difficulties associated with resolving the resonant angle. This is explained in more detail in App.~\ref{sec:laminar}. Certain
other simulations (e.g. $\mu$0.01-$\acceldl$1-$\grainsizedl$0.1), which are at mid- or high-$k$ but with subsonic dust drift ($\langle \driftvelmag\rangle<\cs$),  remain laminar because the RDI is stable, or has very low growth rates, at these parameters.

%
%
\begin{figure*}
\begin{center}
\includegraphics[width=1.0\textwidth]{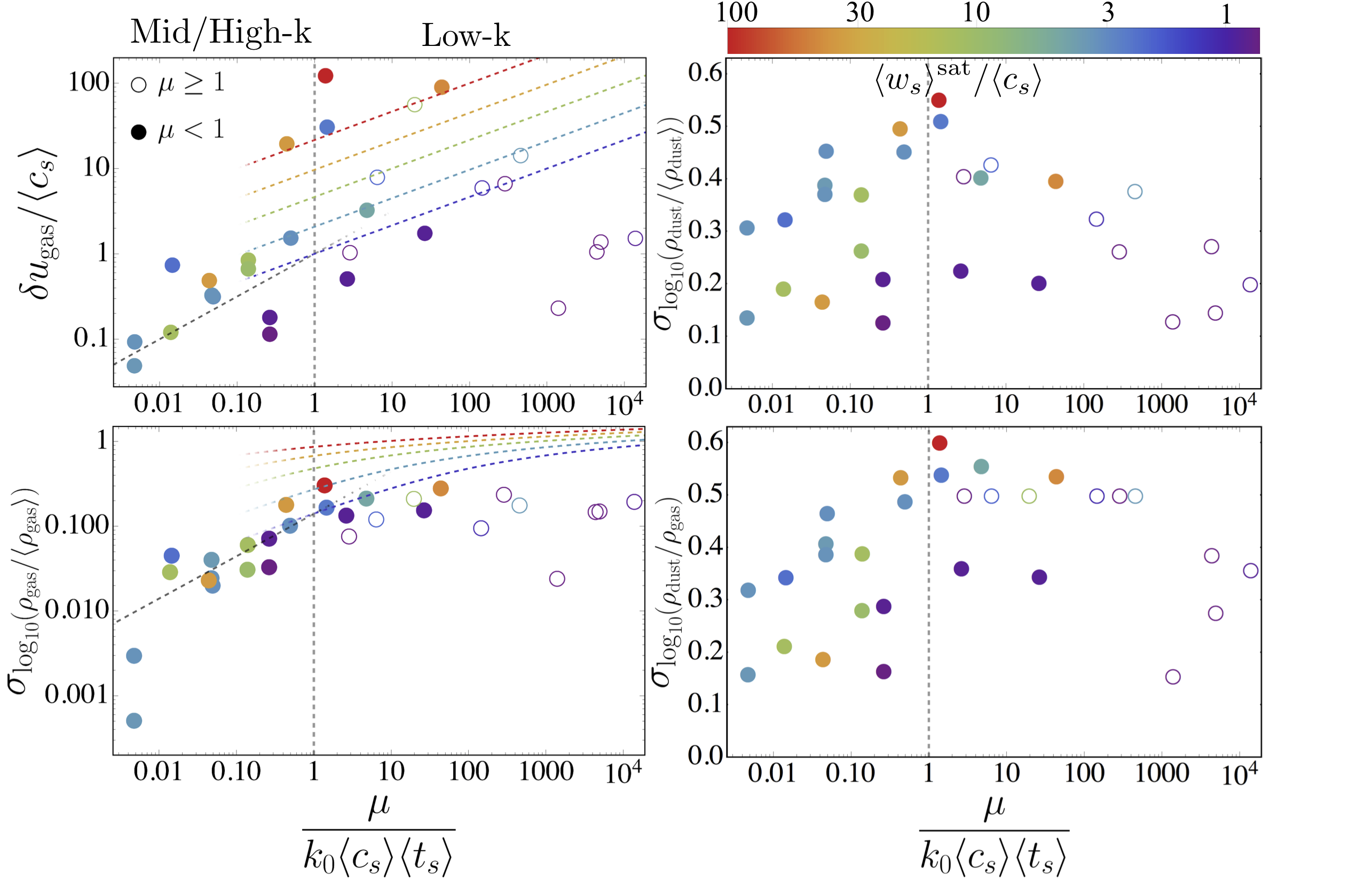}\vspace{-0.5cm}
\caption{Standard deviation of gas velocity, (log) gas density, dust density, and dust-to-gas mass ratio vs.\ $\magicparameter$. Note that  $\magicparameter$ divides the mid/low-$k$ regimes at unity, and the mid/high-$k$ regimes at $\magicparameter=\mu^{2}$.
{\em Top Left:} Dotted lines compare the quasi-linear theory prediction (\S~\ref{sec:theory}), which provides a remarkably good fit except when $\mu \gg 1$ (where several of the assumptions there break down). Note that the prediction depends on $\langle \driftvelmag \rangle/\langle \cs \rangle$ in the low-$k$ regime (hence separate lines for different values of $\langle \driftvelmag \rangle_{\rm sat}/\langle \cs \rangle$, colored as labeled), but does not in the mid/high-$k$ regime, a feature that also appears to be present in the simulation data for $\langle\delta u^2_{\rm gas}\rangle_{\rm sat}^{1/2}$. Large velocity fluctuations are produced at low-$k$ and large $\langle \driftvelmag \rangle/\langle \cs \rangle$. The dust velocity scales similarly, per Fig.~\ref{fig:du}. 
{\em Bottom Left:} Dotted lines show $\sigma(\log_{10}(\rho_{\rm gas}))$, calculated using the predicted $\delta u_{\rm gas}$ (see Fig.~\ref{fig:density_scaling}). Larger velocity fluctuations correlate with larger density fluctuations. 
{\em Top Right:} Dust density fluctuations. There is no obvious predictive relation for the pressure-free dust-density fluctuations.
{\em Bottom Right:} Dust-to-gas-ratio fluctuations. These largely trace the $\dustden$ fluctuations, but are weaker at low-$k$ than would be expected in the case of dust and gas being strictly uncorrelated. That is to say, we observe $\sigma(\log_{10}(\rho_{\rm dust}/\rho_{\rm gas}))^{2} < \sigma(\log_{10}(\rho_{\rm dust}/\langle\rho_{\rm dust}\rangle))^{2}+\sigma(\log_{10}(\rho_{\rm gas}/\langle \rho_{\rm gas}\rangle ))^{2}$ at low-$k$ due to some non-zero covariance between dust and gas density.}
\label{fig:scaling}
\end{center}
\end{figure*}
%
%

\subsubsection{Dust and gas distributions}

Figures~\ref{fig:jointhist} and \ref{fig:pdf} show the distributions of dust and gas (time-averaged in the saturated state). In the low-$k$ regime, dust and gas densities remain broadly proportional to one another in saturated state, even as they vary over $\sim 1-2$\,dex together (albeit with a $1\sigma$ dispersion of $\sim 0.3-0.6\,$dex at fixed $\gasden$). 
This agrees with our expectation that the gas is effectively highly-compressible and moves with the dust at these wavenumbers. As discussed in \citet{hopkins2016fundamentally}, it is well-known that the standard numerical method of integrating trajectories of a finite number of ``super-particles'' can produce some numerical or ``sampling'' noise in the ratio $\dustden/\gasden$; there we show this is at the level of $\sim 0.05$\,dex even when the gas is externally stirred with $\dug/\cs\sim10$ (the noise is still smaller for sub-sonic flows). This is negligible for any simulation here which develops non-laminar behavior (note the effects on $\dug$ and $\sigma(\log_{10}(\gasden))$ are much smaller still, $< 0.01\,$dex).

In contrast, at mid and high-$k$, the gas is largely incompressible (with a small range in $\gasden$, as expected from our arguments above), while the dust occupies a range of densities remarkably similar to the low-$k$ case. This results in larger dust-to-gas ratio fluctuations. {At the same time, low gas density, rarefied regions in the mid/high-$k$ regimes are typically evacuated of dust. This is evidenced by the skewing of the low density tail in the right two panels of Fig.~\ref{fig:jointhist}.}

Note that in simulations that lie near the ``border'' between the low-$k$ and mid-$k$ regimes, the joint gas-dust PDFs show the skewed shape of the incompressible high-$k$ PDF, but with some elongation in $\gasden$ correlated with $\dustden$ as in the low-$k$ PDF.

The one-dimensional PDFs of $\dustden/\gasden$ are shown in Fig.~\ref{fig:pdf}. {In the mid-$k$ and high-$k$ regimes, these PDFs are similar in shape to the PDF of $\dustden$ because fluctuations in $\dustden$ are larger than those in $\gasden$. Because the gas turbulence is more vigorous in the low-$k$ regime, there is only a weak trend towards wider distributions at lower $k$ (higher $\mu/(k_{0}\,\langle c_{s} \rangle \langle t_{s} \rangle)$).
The distributions are very crudely of a  lognormal shape, although many have strong high-$\dustden/\gasden$ tails.}

\subsubsection{Comparison to ``Passive Dust'' Simulations}

Comparing the bivariate PDFs in Figure~\ref{fig:jointhist} to those for simulations of ``passive'' dust clustering in externally-driven isothermal turbulence in \citet{hopkins2016fundamentally}\footnote{Note that \citet{hopkins2016fundamentally} used the same simulation code/numerical methods, and analysis methods.}
(see also \citealt{Pan2013,hogan1999scaling,lee:dynamics.charged.dust.gmcs}), it is clear that the PDFs for the mid-$k$ and high-$k$ regimes here are  {qualitatively} different from the PDFs in those ``passive dust'' experiments, at any turbulent Mach number. 

Even the low-$k$ PDF, which broadly resembles some ``passive dust'' cases at higher Mach number, differs qualitatively in detail. For example, in passive dust simulations (see, e.g. figure 9 of \citealt{hopkins2016fundamentally}), the bivariate PDF always tapers noticeably to $\dustden \approx \mu\,\gasden$ as $\gasden$ increases, because the dust is more  tightly coupled to the gas in dense regions. Here, as seen in the left-hand panel of Fig.~\ref{fig:jointhist}, {this does not occur.} This is because the RDI operates on all density scales and it is the dust that {\em drives} the large density fluctuations in the first place. 

Moreover, the global, strong anisotropy of  RDI turbulence is clear 
in all regimes in Fig.~\ref{fig:regimes}, and is quite different from externally-driven turbulence. Specifically, the dust is arranged into filaments with a preferred direction along the characteristic mode angles of the RDI, even in the nonlinear, turbulent state. Likewise, the dust morphology is quite different, with ``plumes'' and ``jets'' of dust as opposed to ridge-lines between gas vortices.

We have also confirmed these conclusions directly by running a simulation with parameters otherwise identical to an RDI simulation, but with ``passive'' dust ($\mu \rightarrow 0$), and externally-driven turbulence  tuned to produce the same $\dug/\cs$ as the RDI run (using the turbulent-driving scheme of \citealt{hopkins2016fundamentally}). As expected, the bivariate PDFs, morphology, anisotropy, and other key properties have almost nothing in common.  

Not surprisingly, higher-order diagnostics (e.g.\ dust-dust and dust-gas correlation functions) differ even more dramatically between RDI- and externally driven turbulence simulations. These will be studied in  detail in future work.

\subsection{Nonlinear/Saturated Scalings}

\begin{figure}
\begin{center}
\includegraphics[width=0.48\textwidth]{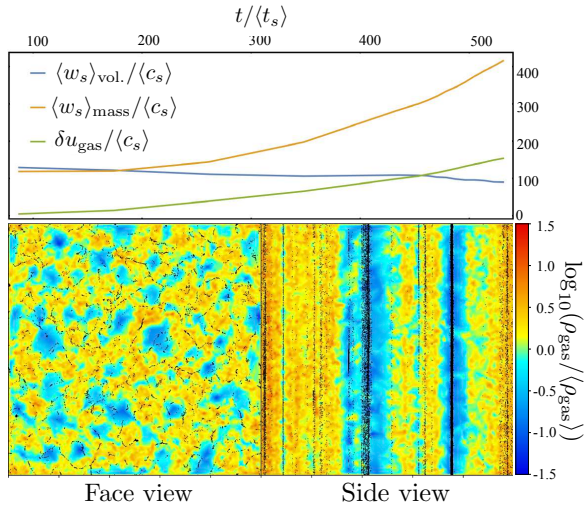}\vspace{-0.3cm}
\caption{Dust drafting in simulation $\mu$0.01-$\acceldl$1e5-$\grainsizedl$0.1. {\em Top:} Volume and mass-weighted dust drift velocities ($\langle \driftvelmag \rangle$) and gas velocity dispersion ($\delta u_{\rm gas}$) versus time. {\em Bottom:} Face/side-on slices of the simulation (as Fig.~\ref{fig:regimes}). When $\langle \driftvelmag \rangle$ and $\mu$ are sufficiently large, dust aligns into \addedtextkey{filaments} along the acceleration direction ${\bf a}$, with particles drafting \addedtextkey{(upstream particles accelerating the gas 
to reduce the drag force on downstream particles). This allows the dust and gas to reach very high speed in the filaments relative to the gas and dust outside of the filaments.} These filaments become increasingly concentrated with time. This means the mass-weighted $\langle w_{s}\rangle_{\rm mass}/c_{s}$ increases, while the volume-weighted $\langle w_{s}\rangle_{\rm vol.}/c_{s}$ remains relatively constant. The gas $\delta u_{\rm gas}$ also continues to grow, but this is increasingly driven by the small volume of gas being entrained in the filaments being rapidly accelerated by high-density dust.}
\label{fig:drafting}
\end{center}
\end{figure}

{A simple diagnostic of the gas turbulence in the saturated state is shown in Fig.~\ref{fig:density_scaling}, 
which plots the gas density dispersion versus its velocity dispersion for 
all simulations, comparing to the expected statistics from driven turbulence simulations Eq.~\eqref{eq:dens.var.mach.num}. We see a reasonable fit for lower Mach numbers,
indicating the gas turbulence is broadly similar to standard (solenoidally) forced turbulence. There are, however, significant differences
once $\delta u_{\rm gas}\gtrsim 3 c_s$., which is likely because in this high-Mach-number regime the system becomes highly anisotropic, with strong channels of high-velocity gas driven by dust columns (see \S\ref{sec:drafting} and Fig.~\ref{fig:drafting}).}

Figure~\ref{fig:du} compares the saturated dust and gas velocity dispersions. As expected from equipartition arguments, $\dud\approx\dug$ in nearly all simulations.

Figure~\ref{fig:scaling} shows the saturated values of $\dug/\cs$, for our full set of simulations. We compare to the predicted mid-$k$ and low-$k$ scalings from \S~\ref{sec:theory} (Eq.~\ref{eq:duscaling}), which do a surprisingly good job explaining the non-linear behavior, at least at the order-of-magnitude level. Performing a maximum likelihood fit to the data in each regime, modeling $\dug/\cs \propto \mu^{\alpha_{\mu}} (\wavenumber)^{\alpha_{k}} (\langle \driftvelmag \rangle/\cs)^{\alpha_{w}}$, we find ``best-fit'' power-laws consistent with our predicted scaling within the $\sim 1\sigma$ range, but with large uncertainties arising from the limited statistics.

It is worth noting that the mass-weighted statistics conform better to the scalings, but for consistency with the rest of the figures we have chosen to display volume-weighted statistics. Our low-$k$ scalings, while holding broadly, have a large scatter for  $\dug/\cs$. The predictions for $\sigma[\log_{10}( \gasden /\langle \gasden \rangle)]$  are somewhat too high at low-$k$, although they are more accurate for the mid- and high-$k$ regimes. This is reflective of an apparent plateau in $\sigma[\log_{10}(\gasden/\langle \gasden \rangle)]$ at $\sigma[\log_{10}(\gasden/\langle \gasden \rangle)]\sim 0.2$ (see also Fig.~\ref{fig:density_scaling}).

We did not have an {\em a priori} prediction for the magnitude of dust density fluctuations ($\sigma[\log_{10}(\dustden)]$ or $\sigma[\log_{10}(\dustden/\gasden)]$, which are similar), but Figure~\ref{fig:scaling} shows they increase both with $\mu/(\wavenumber)$ and $\langle \driftvelmag \rangle/\cs$, reaching a maximum of $\sim 0.5-0.6$\,dex dispersion.

%
%
%

%

\subsection{Extreme Cases: Decoupling of Dust ``Jets''}
\label{sec:drafting}

In extreme, high-$\driftvelmag$ simulations, we observe a phenomenon we term ``dust drafting,'' where  dust aligns into narrow filaments in the direction of motion.
\addedtextkey{These filaments drag a small fraction of the gas along with them, leaving the rest of the gas behind. A minority of the gas gets a majority of the force, $\mu {\bf a}$, causing
these filaments effectively decouple from the rest of the gas (cf. Eq.~\ref{eq:tsdrift}).}

Figure~\ref{fig:drafting} shows one example: {\small $\mu$0.01-$\acceldl$1e5-$\grainsizedl$0.1}, with  $w_{s,{\rm eq}}/\cs \sim 100$, the highest in all our simulations. 
\addedtextkey{ A small amount of gas remains tightly coupled to the dust and is dragged along with it, such that \textit{within} the filaments, the drift velocity $\driftvelmag$ is closer to its equilibrium value. This is reflected in Figure ~\ref{fig:drafting}'s upper panel, where the volume-weighted drift velocity (termed $\langle \driftvelmag \rangle _{\rm vol.}$) differs from the mass-weighted drift velocity (termed $\langle \driftvelmag \rangle _{\rm mass}$) by more than a factor of four.}
 We have observed this phenomenon only for simulations with 
$\driftveleq/\cs \gtrsim 10$, and the onset is more rapid at higher $\mu$.

\addedtextkey{In our simulations, once a filament forms it continues accelerating indefinitely, although in reality there could be stronger Kelvin-Helmholtz instabilities or viscous dissipation induced by the strong shear at the boundary of the filament (because the width of the filament becomes very small and may depend on resolution, such effects are difficult to resolve numerically).}
The mass-weighted $\langle \driftvelmag \rangle$ continues to grow, reaching $\langle \driftvelmag \rangle_{\rm mass}/\cs\sim 400$ at the time shown, while the volume-weighted $\langle \driftvelmag \rangle$ is approximately constant. The density statistics remain stable as this happens, while $\dug/\cs$ continues to marginally increase.

Of course, this phenomenon is partly an artifact of using a finite, periodic simulation domain. In a global simulation with more realistic dust physics, these filaments might be driven out of the gas altogether, or trigger the onset of secondary effects like dust self-shielding and/or grain collisions.

\subsection{Effects of the Gas Equation-of-State}

Our default simulations adopt an isothermal equation-of-state (EOS) for the reasons discussed above (\S~\ref{sec:methods}). However we have also re-run the simulation $\mu$0.01-$\acceldl$1e3-$\grainsizedl$0.1 
using a strictly polytropic (constant-entropy) EOS with $P = \gamma^{-1}\,\langle \cs \rangle^{2}\,\langle \gasden \rangle\,(\gasden/\langle \gasden \rangle)^\gamma$, with $\gamma=5/3$ (this is labeled with the suffix "-$\gamma$5/3" in Tab.~\ref{tab:sims}). Note that we effectively assume that the gas reverts instantaneously to the polytropic EOS  after shocks, rather than allowing the entropy to increase (otherwise the box-averaged pressure and sound speed would increase in time without limit). In the linear regime, the different EOS  modifies only an order-unity pre-factor in the RDI growth rate
and  we find qualitatively similar results in the simulation here. 

In the saturated state, we find the gas turbulence and resulting gas density fluctuations are modestly suppressed for the stiffer EOS, as expected, but this is again an order-unity effect (note that turbulent Mach numbers are sub-sonic here). Further, the PDF of $\gasden/\dustden$ is narrower in the $\gamma = 5/3$ case than in the $\gamma = 1$.

\section{Conclusions}
\label{sec:conclusions}

In this paper, we present the first study of the non-linear regime of the acoustic RDI. We focus on the simple case of dust grains in a homogeneous medium, coupled via aerodynamic (Epstein) drag, with a constant differential acceleration between gas and dust. We find that the acoustic RDI grows robustly at all scales, eventually breaking up into internally generated turbulence and saturating at large amplitudes. The turbulence is highly anisotropic, with dust concentrated in filaments, plumes or jets along the direction of acceleration. Strongly non-linear structures occur consistently in the dust, necessitating numerical simulations that can follow the velocity distribution function of the grains. We show the simulations can be conveniently characterized by three dimensionless numbers, and survey the parameter space to characterize the saturated states. 

The linear growth rates and structure (e.g.\ wavenumbers and their ``resonant angles'') of the fastest-growing modes agree with the predictions from linear theory \citep{hopkins2018resonant}. The resonant angles are sufficiently virulent to persist well into the non-linear state, and can be clearly seen in the turbulence. The behavior in the linear regime can be organized into three regimes based on the range of wavenumber $\wavenumber$ (``low,'' ``mid,'' and ``high''-$k$), and we find this division persists in the non-linear regime. 

The mid and high-$k$ regimes seem qualitatively similar (but quantitatively distinct): the turbulence driven in the gas is sub-sonic, and only weakly compressible. Although the dust and gas local velocity dispersions reach rough equipartition in saturation, the dust density structure is strongly modified, with the aforementioned plumes and filaments appearing. This generates a distribution of $\dustden/\gasden$ which has an approximate log-normal (or perhaps power-law-like in the tails) shape, with $1\sigma$ dispersions reaching $\sim 0.3-0.6\,$dex (reaching orders-of-magnitude fluctuations, in the tails).

The low-$k$ regime is essentially defined by wavenumbers where the pressure gradient forces in equilibrium are weak compared to the bulk force exerted by dust on gas. This makes the gas more compressible, and it is driven by the dust into large density fluctuations, and even shocks. To leading order dust and gas densities are correlated, albeit with significant scatter at any $\gasden$. The fastest-growing mode structures are distinct, with wavenumber aligned with the acceleration direction, producing long-wavelength ``arcs'' or ``shells'' in dust and gas. {We see broadly similar behavior in the dust-dominated regime with $\dustden/\gasden\geq 1$, consistent with the idea that the character of the instability is similar in this regime
\citep{hopkins2018resonant}.}

In all cases, we note that the non-linear behavior is qualitatively different from the results of ``passive'' dust simulations, in which the dust is treated as a pure tracer population (i.e.\ the gas does not ``feel'' the dust), with externally-driven turbulence.

These simulations form a first step towards studying the effects of the acoustic RDI on astrophysical phenomena. Certainly, the large dust density and velocity fluctuations and non-linear, anisotropic concentrations produced, will alter critical properties such as dust-gas chemistry, dust growth and collision rates, extinction and effective/observable attenuation curves, and more. Future work will consider detailed applications to specific astrophysical environments -- e.g.\ cool-star winds, AGN torii, and dense GMCs -- but these require additional physics (e.g.\ radiative cooling, or global simulations with outflow boundaries) that break the scale-free nature of our studies here. While we work in dimensionless problem units here, in \citet{hopkins2018resonant} we discuss how these translate to physical units in the astrophysical environments mentioned above. For reference, the boundary between low- and mid-$k$ regimes (long and mid-wavelength modes) occurs around scales of $\sim\!0.001-0.1\,$pc in GMCs, $\sim \!1-100$\,au in AGN torii, and $\sim \!10^{3}-10^{5}\,$km in cool-star wind environments, for typical parameters. In all cases, these are  relatively small scales compared to those of the systems in question.

In this paper, in order to aid physical understanding, we simplified  by assuming a single grain size in each simulation. This implies that quantities like the stopping time, resonant angle, and drift velocity are single-valued. In future work, we will explore the more physical case with a spectrum of grain sizes. We will also explore higher-order diagnostics, e.g.\ grain clustering and collision kernels. Finally, a wide variety of other RDIs remain to be explored in the nonlinear regime, with many (e.g.\  magnetohydrodynamical RDIs, or the epicyclic RDI) likely to have qualitatively different non-linear behavior and and saturation mechanisms.

\begin{small}\section*{Acknowledgments}\end{small}

We would like to thank Stefania Moroianu for a number of helpful discussions. Support for PFH, JS, \&\ ERM was provided by an Alfred P. Sloan Research Fellowship, NSF Collaborative Research Grant \#1715847 and CAREER grant \#1455342, and NASA grants NNX15AT06G, JPL 1589742, 17-ATP17-0214. JS was also supported by the Royal Society Te Ap\=arangi through contracts RDF-U001804 and UOO1727.  Numerical calculations were run on the Caltech compute cluster ``Wheeler,'' allocations from XSEDE TG-AST130039 and PRAC NSF.1713353 supported by the NSF, and NASA HEC SMD-16-7592.

\vspace{0.2cm}

\bibliographystyle{mnras}
\bibliography{ms_reduced}

\appendix

\section{Numerical Limitations in the High-Wavenumber Regime}
\label{sec:laminar}

In some simulations -- in particular those at high dimensionless wavenumber, $\wavenumber$ -- the RDI does not grow as expected from  linear-theory, with the gas remaining laminar (or nearly so) over the duration of the simulation.  We do not believe this to be a physical effect, but rather due to finite resolution and numerical dissipation/noise. We discuss the possible causes for this here, in order to motivate further study of this regime. The affected simulations are shown with gray text in Table~\ref{tab:sims} (recall that there are also some simulations, e.g. $\mu$0.01-$\acceldl$1-$\grainsizedl$0.1, that do not go turbulent because they are subsonic and in the mid-$k$ or high-$k$ regime, where growth rates are very low; see \citealt{hopkins2018resonant}).

A particular challenge for simulating the RDI in the high $\wavenumber$ regime is resolving the resonant angle. While we can always construct a box that resolves arbitrarily high  $\wavenumber$ by changing the simulation parameters ($\acceldl$ and $\grainsizedl$), as shown in \citet{hopkins2018resonant}, the linear growth  rates become increasingly sharply peaked around the resonant angle (where $\hat{\bf k}\cdot \driftveleq = c_s$) with increasing $\wavenumber$. More precisely, defining the {longest wavelength} mode in the box as 
$k_{0} = \xi/(\mu\,\cs\,\langle\ts\rangle)$ (where $\xi > 1$ is required to be in the high-$k$ regime), the growth rate  drops by orders of magnitude outside some width in mode angle, $\Delta\cos{\theta_k} \sim \Delta \theta_k \sim \mu\,\xi^{-1/3}\,(\langle\driftvelmag\rangle/\cs)^{-1}$ (in the high-$k$ regime, {assuming $\langle w_s \rangle \gg c_s $}). At finite resolution, any angular structure smaller than $\Delta \theta \sim \Delta x / L_{0} = N^{-1/3}$ will be unable to be resolved. At our fiducial resolution of $N^{1/3}=128$, this means we can only barely resolve resonant angles in the high-$k$ regime for marginally high-$k$ boxes ($\xi\sim1$), relatively low $\langle\driftvelmag\rangle/\cs \sim 1-10$, and relatively high $\mu\sim 0.1$.  Examination of the parameters in Table~\ref{tab:sims} shows that almost all  simulations that fail to become turbulent do indeed have very narrow RDI resonances. Unfortunately, this numerical issue is difficult to overcome (e.g. fixed-grid-based codes almost certainly face the same issues) and even significant increases in resolution allow only modest gains in the resolvable resonant angles (and thus the $\wavenumber$ that is possible to simulate). Two-dimensional simulations, which allow much higher resolutions, could thus be particularly helpful for study of the high-$k$ regime.

There is another numerical  issue, again primarily affecting the high-$k$ regime, which is more specific to the Lagrangian finite-volume method used here. 
In the linear regime, at increasing $k$, the back-reaction from drag becomes an increasingly small perturbation compared to pressure forces (as $\nabla P \sim k\,P$). Non-linearly, as discussed in \S~\ref{sec:theory}, this translates to the predicted $\dug/\cs$: again defining $k_{0} = \xi/(\mu\,\cs\,\langle\ts\rangle)$,  then in the high-$k$ regime, we see that $\dug/\cs \sim \mu\,\xi^{-2/3}$, which is very small (see Eq.~\ref{eq:duscaling}). It is well-known that both finite-volume and Lagrangian numerical hydrodynamics methods have difficulty accurately capturing very low Mach number, nearly-incompressible turbulence: the Riemann solver introduces numerical diffusion and the constant re-arrangement of the grid introduces ``remeshing noise,'' which launches sound waves, both of which make it difficult to follow highly-subsonic effects. In \citet{hopkins2014gizmo}, we show specifically for the numerical methods here that sub-sonic effects can be numerically over-damped below turbulent Mach numbers $\ll 0.01$. 

To capture RDI-induced turbulence at such low $\dug/\cs$, other, lower-noise, numerical methods
may be required.
Note that in every case tested where the instabilities failed to grow, either the predicted  $\dug/\cs \ll 0.01$, or the resonant angle was unresolved  (usually both).

\section{Additional Numerical Details \&\ Tests}
\label{numerics}

%
%
\begin{figure}
\begin{center}
\includegraphics[width=1\columnwidth]{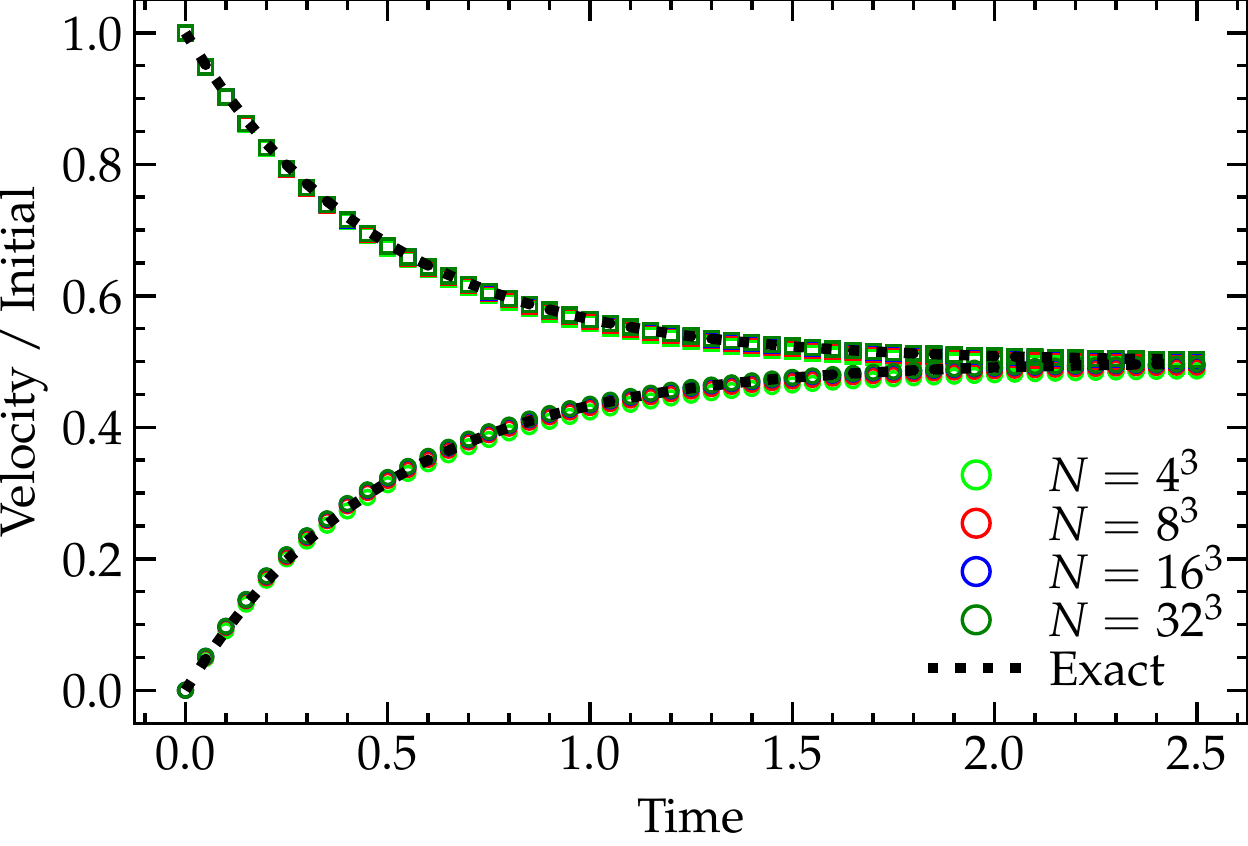}
\includegraphics[width=1\columnwidth]{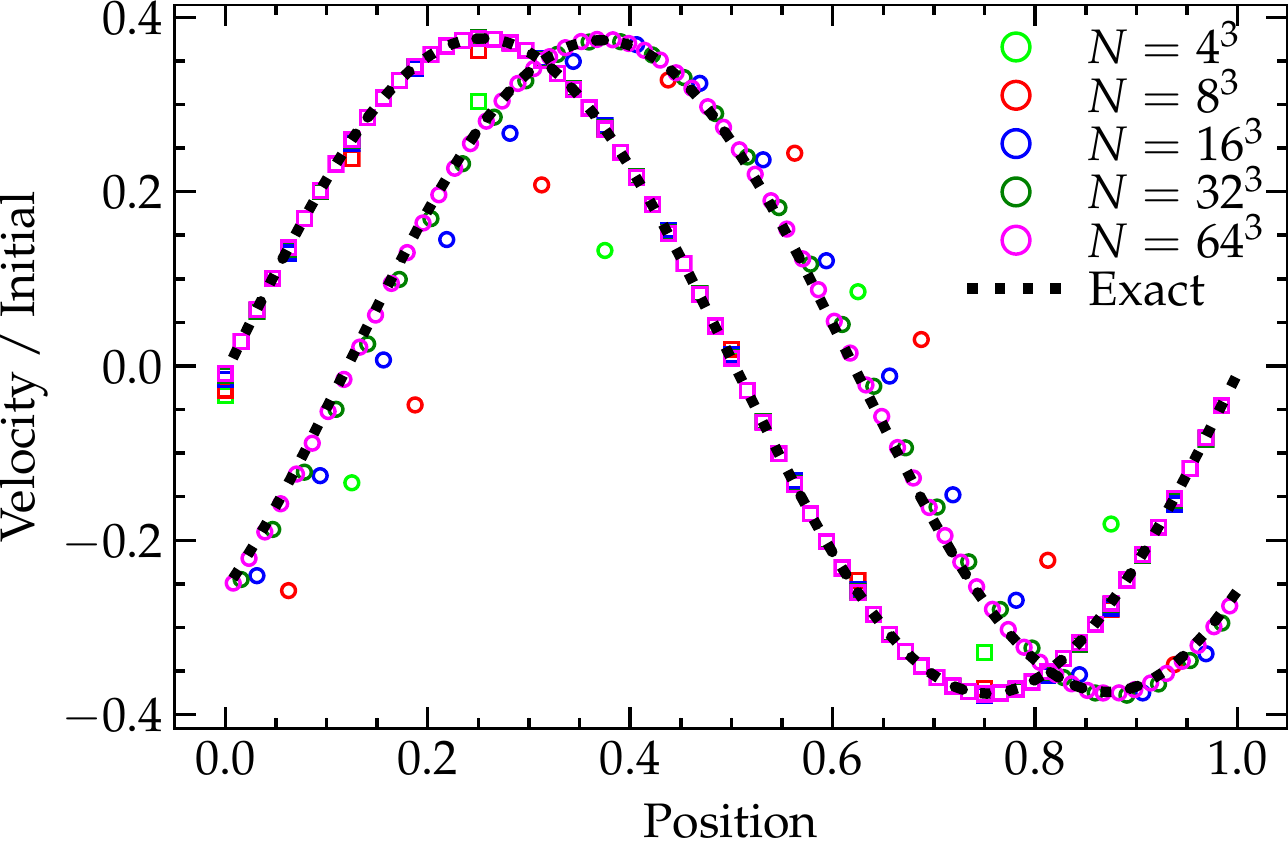}
\vspace{-0.5cm}
\caption{Validation tests of the dust-gas coupling algorithm used in the main text. {\em Top:} Uniform box of dust and gas (no external acceleration) with non-zero initial $\dustvel$ (but $\gasvel=0$). We plot the box-averaged velocity versus time (in code units), for both gas (circles) and dust (squares), for 3D boxes with different resolution (labeled). Convergence to the exact solution is rapid (at time $1.5$, the L1 error norm at $N=4^{3}$ [$16^{3}$] is $\sim 0.01$ [$\sim 0.001$]). {\em Bottom:} Damped, coupled linear dust-gas acoustic waves, at time $t=1.2\,\lambda/c_{s}$. Convergence in the {\em dust} wave (squares) is extremely rapid. The {\em gas} wave (circles) converges more slowly owing to its sensitivity to the ``smearing'' of the momentum ``back-reaction'' force from dust onto gas (which is only first-order). Still, even at $N=4^{3}$, all qualitative features of the waves are captured, and by $N=32^{3}$ the deviations from the exact solution are nearly indistinguishable.}
\label{fig:dustyboxwave}
\end{center}
\end{figure}
%
%

\subsection{Numerical Implementation}
 As noted in the main text, the salient equations \eqref{eq:eom}--\eqref{eq:euler} are solved using the code {\small GIZMO} \citep{hopkins2014gizmo}. The numerical scheme for hydrodynamics and magneto-hydrodynamics (i.e.\ the gas equations, absent dust) in {\small GIZMO} has been extensively described and tested in previous work \citep[e.g.][]{gaburov:2011.meshless.dg.particle.method,hopkins2014gizmo,hopkins:mhd.gizmo,hopkins:cg.mhd.gizmo,hopkins:gizmo.diffusion,zhu:2016.sph.vs.gizmo.cosmo.sims.mw.mass.galaxy,2017ApJ...847...43D,2018MNRAS.473.1603H}. Likewise, the scheme for integrating the trajectories of dust particles within gas is described and validated in detail in previous studies \citep[e.g.][]{hopkins2016fundamentally,lee:dynamics.charged.dust.gmcs}.

The only added numerical element in {\small GIZMO} in this study is the ``back-reaction'' term to the gas equation-of-motion, i.e.\ the force from dust on gas. Our implementation follows standard well-tested methods from e.g.\ \citet{2007ApJ...662..613Y,bai:2010.grain.streaming.sims.test}: after calculating the change to the velocity/momentum of a given dust super-particle $\Delta {\bf p}_{a}$ (which is integrated semi-implicitly over the entire timestep, as described in \citealt{hopkins2016fundamentally}), the momentum is subtracted from the surrounding gas elements according to the weighted kernel function $\Delta {\bf p}_{b} = - \Delta {\bf p}_{a}\,W({\bf x}_{b}-{\bf x}_{a},\,H_{a}) / \sum_{c} W({\bf x}_{c}-{\bf x}_{a},\,H_{a})$, where $W$ is the  same kernel function used to define both the hydrodynamic operations and the interpolation of gas properties to the grain super-particle position (unlike grid or spectral hydrodynamics methods, since our code is particle based, there is no ambiguity about the appropriate ``matching'' kernel function). In this study we adopt the standard cubic spline for $W$, with radius of compact support $H_{a}$ set to twice the kernel-averaged gas element neighbor distance (this is identical to the hydrodynamic search, see \citealt{hopkins2014gizmo}). The normalization of $W $ ensures total momentum conservation is machine-accurate, and because our hydrodynamic method is finite-volume, the momentum flux $\Delta{\bf p}_{b}$ is treated like any other hydrodynamic flux in the drift-kick operations and timestep restrictions for gas.

In Fig.~\ref{fig:dustyboxwave}, we consider two common numerical validation tests for coupled dust-gas dynamics, variants of the ``dustybox'' and ``dustywave'' problems from \citet{2011MNRAS.418.1491L}. In both, we initialize a homogeneous 3D periodic box (size unity) with mean $\langle\dustden\rangle=\langle\gasden\rangle=c_{s} = L_{\rm box}=t_{s}=1$. In the first, we set $\gasvel=0$, $\dustvel=v_{0}\,\hat{x}$; this has a trivial analytic solution where the dust decelerates, while accelerating the gas, until the two reach the same velocity $=v_{0}/2$. In the second, we initialize an adiabatic traveling coupled linear dust-gas wave with $\delta v/c_{s} = 2\,\delta\rho/\rho = 10^{-4}\,\sin(2\pi\,x)$ in dust and gas; this corresponds to a coupled dust-gas wave system where the two waves come in and out of phase while gradually damping (analytic solutions here are less trivial, but are described in detail in \citealt{2011MNRAS.418.1491L}). Both exhibit the expected behavior and convergence rates.

Similar algorithms have been studied extensively in the literature (for some examples in astrophysical applications, see e.g.\ \citealt{2007ApJ...662..613Y,bai:2010.grain.streaming.sims.test}, or for examples of widespread applications in laboratory/terrestrial particle-laden turbulence, see \citealt{1988JCoPh..79..373Y,1993PhFlA...5.1790E,1994JFM...277..109K,1996PhFl....8.2733P,1997JFM...335...75S,1998JFM...375..235B,1999JFM...379..105S,2003PhFl...15..315F,2010JFM...650....5L,2017JCoPh.338..405I}). For readers interested in the exact numerical implementation details (because it is impossible to be complete in describing all aspects of the code parallelization and other aspects), we make our methods public and distribute them directly in the public {\small GIZMO} source code, alongside detailed User Guide and test problem setups.\footnote{See \href{http://www.tapir.caltech.edu/~phopkins/Site/GIZMO.html}{\url{http://www.tapir.caltech.edu/~phopkins/Site/GIZMO.html}}} We stress that this includes all algorithms used here and can reproduce all our results in this paper.

\subsection{Robustness to Numerical Methods \&\ Initial Conditions}

%
%
\begin{figure}
    \centering
    \includegraphics[width=1\columnwidth]{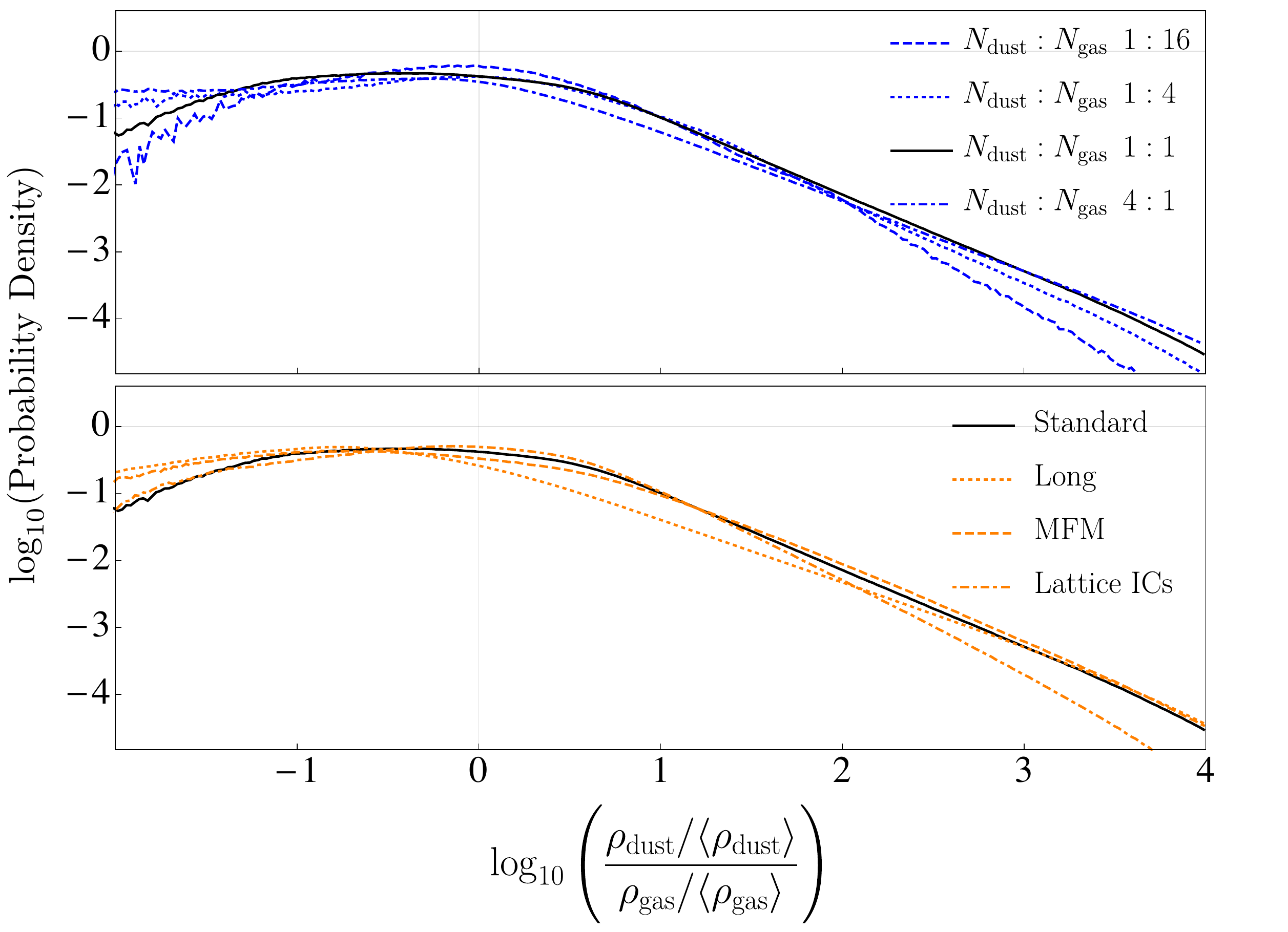}
    \vspace{-0.5cm}
    \caption{Various 2D numerical tests validating that our methods are robust for study of the RDI. All have $256^{2}$ gas particles and, other than being two dimensional, are share parameters with the simulation $\midk$ (cf. Tab.~\ref{tab:sims}). These PDFs are computed identically to those in Fig.~\ref{fig:variance}. {\em Top:} These simulations are identical except for a varying number of dust particles. The legend shows the ratio of dust to gas particles, with the lowest number being just 1/16 as many dust particles as gas particles, and the highest being 4 times as many. The PDF does not notably change aside from becoming noisier as we use fewer dust particles. {\em Bottom:} Depicted here are the results from doubling the box length (Long), using an alternative numerical method (meshless finite-mass (MFM)), and starting from lattice-like initial conditions, rather than glass-like (Lattice ICs). All PDFs are again quite similar. Statistical fluctuations likely account for any discrepancies. 
    }
    \label{fig:NumTests}
\end{figure}
%
%

%
\begin{figure}
\begin{center}
\includegraphics[width=0.5\textwidth]{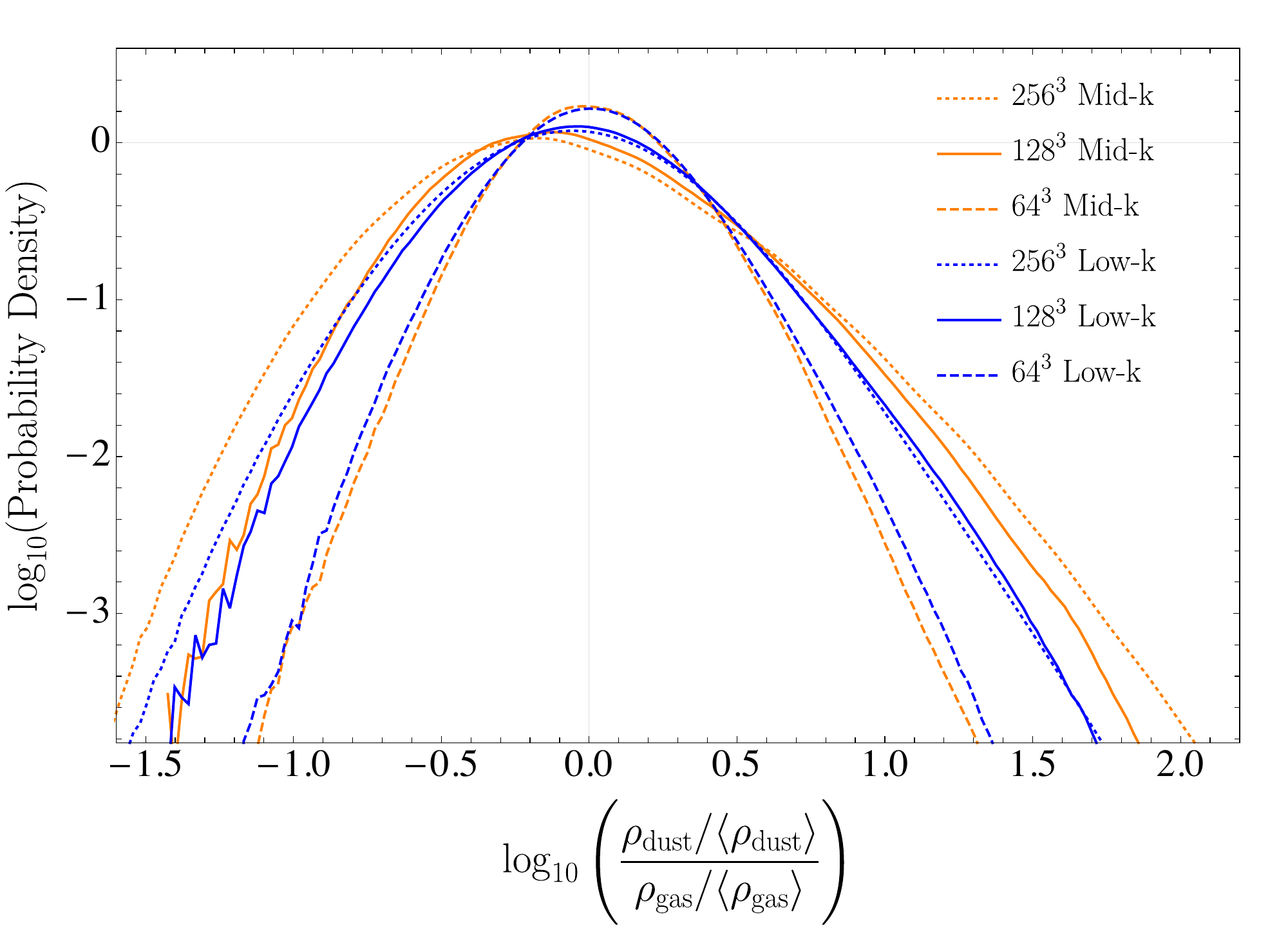}\vspace{-0.3cm}
\caption{The volume weighted PDFs of dust density over gas density for the simulations \lowk {(low-$k$)}, \midk {(mid-$k$)}\  and their high/low resolution counterparts $\mu$0.01-$\acceldl$1e4-$\grainsizedl$0.001-HR/LR and $\mu$0.01-$\acceldl$100-$\grainsizedl$0.1-HR/LR. In the majority of our simulations, we have $128^{3}$ dust and gas particles. In order to test the effects of resolution on our results, we have run and examined two simulations with $256^3$ dust and gas particles and two with $64^{3}$. The particular parameters of these simulations (i.e. $\mu$, $\bar{\epsilon}$, and $\bar{a}$) are identical to those of \lowk and \midk. We find that our results seem relatively well converged at $128^3$. The low-density tails of the PDFs here are not sampled quite as well in the $128^{3}$ simulations as in the $256^3$, while on the high-density end the PDFs are better resolved. In the low resolution, $64^3$ simulations, the high and low density regions are  more poorly sampled, and the simulations fail to accurately capture the non-Gaussian structure of the PDFs.  Other statistics associated with the saturated state of the simulations are also mostly unchanged when increasing the resolution (c.f. Table \ref{tab:sims}).
}
\label{fig:variance}
\end{center}
\end{figure}
%
%
%
Unfortunately, the acoustic RDI studied here is not amenable to idealized ``validation'' tests, nor is there a well-defined convergence criterion or error norm. Exact analytic non-linear solutions or any other ``reference solution'' against which to compare do not exist, hence our motivation for this study. Even in the linear regime, the problem fundamentally is that {\em all} wavelengths are unstable with a growth rate that increases {\em monotonically and without limit} with decreasing wavelength.\footnote{Consider for example running a linear-regime test problem with a ``seeded'' mode in the low-$k$ regime at wavelength $\lambda = 5\,\Delta x_{0}$ ($\Delta x_{0}$ the initial resolution scale), run for just one $e$-folding time. If we surveyed a factor of just $\sim 32$ in linear resolution, then grid-scale modes (seeded e.g.\ by integration error) in the highest-resolution case should (according to the extrapolation of linear theory) grow in amplitude by a factor of $\sim 10^{13}$ in the same time.} 
In other words, the ``fastest-growing mode'' is always at the grid-scale, and we {\em should} expect to obtain different solutions at different resolution (at any time).

This is a uniquely challenging aspect of these studies, which merits further exploration in future work. For now, however, we can ask a simpler question: whether the bulk properties of the boxes simulated are especially sensitive to the details of the numerical method or resolution adopted (over the range of what is computationally feasible).

As a test of how robust our statistics are to variations in numerical methods, we have run several different 2D simulations, each with the same parameters as our mid-$k$ case study, $\midk$. PDFs of the ratio of dust density to gas density are shown in Fig.~\ref{fig:NumTests}, averaged over the saturated state. In one simulation, the box is twice as long in the direction of acceleration ${\bf a}$. As expected, there are no notable changes in this case. In another simulation, we have used the second-order Lagrangian finite-mass ``meshless finite-mass'' (MFM) method for the hydrodynamics instead of our preferred ``meshless finite-volume'' (MFV) method. As particles have constant mass in this method, the high-density regions are better sampled, while at the same time the low-density regions are poorly sampled, and noisy. Even so, the results are consistent with the results obtained with MFV. We have also tried initializing gas and dust particles at grid points, rather than random positions. The results obtained are, within statistical fluctuations, identical to the glass-like, randomly sampled initial conditions.
\subsection{Robustness to Numerical Resolution}

In Fig.~\ref{fig:variance}, we illustrate the dependence of the density PDFs on resolution  for the low-$k$ and mid-$k$ simulations, which are not strongly affected by the  numerical difficulties discussed above. We see that the differences between the $128^3$ and $256^3$ simulations are minimal, although, perhaps unsurprisingly, the highest and lowest $\dustden/\gasden$ are somewhat unresolved at  $128^3$. At $64^3$ the PDFs look significantly more Gaussian, illustrating that aspects of the turbulence may not be well-resolved. We note, however, that the general characteristics of the RDI -- e.g.\ the resonant angle and general structure of the instability -- look very similar at all three resolutions (not shown).\\

In addition to studying resolution effects in 3D, we have examined the effects of using various different numbers of dust particles in 2D simulations. PDFs of the dust-to-gas mass ratio are shown in the top panel of Fig.~\ref{fig:NumTests}. {It is important to note that these PDFs differ substantially from the three dimensional case, as should be expected due to the different properties of turbulence in two dimensions.  For this reason, we have chosen to isolate the two dimensional tests rather than present them all together.} Even when using 1/16 as many dust particles as gas particles, the PDF remains, while noisier, relatively similar to when higher numbers of dust particles are used. Thus, including a greater or lesser number of dust particles does not seem to dramatically impact our results.

\end{document}